\documentclass[conference,compsoc]{./IEEEtran}

\usepackage{fontawesome5}
\usepackage{tikz}
\usetikzlibrary{arrows.meta, positioning, calc}
\usetikzlibrary{backgrounds}
\usetikzlibrary{patterns}
\usepackage{amssymb}\usepackage{pifont}
\newcommand{\cmark}{\ding{51}}\usepackage{tabularray}
\usepackage{pgfplots}
\pgfplotsset{compat=1.11}
\usepgfplotslibrary{fillbetween}
\usepackage{tablefootnote}
\usepackage{forest}
\usepackage{float}
\usepackage[hidelinks]{hyperref}

\usepackage{booktabs}
\usepackage{makecell}
\usepackage{tabularx}

\usepackage{threeparttable}
\usepackage{enumitem} \usepackage{xcolor}
\usepackage{sansmath}

\usepackage{acro}

\ifCLASSOPTIONcompsoc
\usepackage[nocompress]{cite}
\else
\usepackage{cite}
\fi

\usepackage{amsmath}
\usepackage{url}

\usepackage{xcolor}
\usepackage{ifthen}

\newcommand{\numcomsintaxonomy}{6}

\newcommand{\fuzzername}{\emph{SyncWordFuzzer}}

\DeclareAcronym{obc}{
  short=OBC,
  long=On-Board Computer
}
\DeclareAcronym{cdhs}{
  short=CDHS,
  long=Command and Data Handling System
}

\DeclareAcronym{adcs}{
  short=ADCS,
  long=Attitude Determination and Control System
}

\DeclareAcronym{com}{
  short=COM,
  long=Communication Subsystem,
}

\DeclareAcronym{ttnc}{
  short=TT\&C,
  long=Telemetry Tracking and Control,
}

\DeclareAcronym{leo}{
  short=LEO,
  long=Low Earth Orbit,
}

\DeclareAcronym{mcs}{
  short=MCS,
  long=Mission Control System,
}
\newcommand{\mcs}{\ac{mcs}}

\DeclareAcronym{tc}{
  short=TC,
  long=Telecommand,
  long-plural = s,
}
\newcommand{\tc}{\ac{tc}}

\DeclareAcronym{tm}{
  short=TM,
  long=Telemetry,
}
\newcommand{\tm}{\ac{tm}}

\DeclareAcronym{eps}{
  short=EPS,
  long=Electrical Power System,
}
\newcommand{\eps}{\ac{eps}}

\DeclareAcronym{fep}{
  short=FEP,
  long=Front End Processor,
  long-plural = s,
}
\newcommand{\fep}{\ac{fep}}

\DeclareAcronym{gs}{
  short=GS,
  long=Ground Station,
}
\newcommand{\gstation}{\ac{gs}}

\newcommand{\etal}{et al.}

\DeclareAcronym{dos}{
  short=DoS,
  long=Denial of Service,
}

\DeclareAcronym{csp}{
  short=CSP,
  long=Cubesat Space Protocol,
}

\newcommand{\leo}{\ac{leo}}

\newcommand{\ttnc}{\ac{ttnc}}

\newcommand{\GomSpace}{GomSpace}
\newcommand{\Endurosat}{Endurosat}
\newcommand{\LibreSpace}{Libre Space Foundation}

\newcommand{\AXonehundred}{NanoCom AX100}

\newcommand{\GSonehundred}{NanoCom GS100}
\newcommand{\UHFtwo}{UHF II}

\newcommand{\NogsComm}{SatNOGS-COMMS}
\newcommand{\satnogs}{SatNOGS}

\newcommand{\cdhs}{\ac{cdhs}}
\newcommand{\com}{\ac{com}}
\newcommand{\ocs}{Orbit Control System}

\newcommand{\adcs}{\ac{adcs}}
\newcommand{\obc}{\ac{obc}}

\newcommand{\coms}{\acp{com}}
\newcommand{\csp}{\ac{csp}}
\newcommand{\esps}{Endurosat Protocol Stack I}

\newcommand{\cots}{COTS}
\newcommand{\satbleed}{\textsc{SatBleed}}
\newcommand{\smallsat}{Small Satellite}

\definecolor{PAcolor}{HTML}{D62728}      \definecolor{AitMcolor}{HTML}{68bbe3}    \definecolor{MTcolor}{HTML}{2CA02C}      \definecolor{GSAcolor}{HTML}{FF7F0E}     \definecolor{MIcolor}{HTML}{9467BD}

\newcommand{\attackerPA}{\colorbox{PAcolor}{\textcolor{white}{\textbf{PA}}}}
\newcommand{\attackerAitM}{\colorbox{AitMcolor}{\textcolor{white}{\textbf{TTCLA}}}}
\newcommand{\attackerMT}{\colorbox{MTcolor}{\textcolor{white}{\textbf{MT}}}}
\newcommand{\attackerGSA}{\colorbox{GSAcolor}{\textcolor{white}{\textbf{GSA}}}}
\newcommand{\attackerMI}{\colorbox{MIcolor}{\textcolor{white}{\textbf{MI}}}}

\newcommand{\foundVuln}{\faFlag}

\begin{document}
\title{\textsc{\satbleed{}}: Security of Commoditized Communication Modules in Satellites}

\author{\IEEEauthorblockN{Ulysse Planta\IEEEauthorrefmark{1},
Julian Rederlechner\IEEEauthorrefmark{1},
Martin Strohmeier\IEEEauthorrefmark{2},
Mathias Fischer\IEEEauthorrefmark{3} and
Ali Abbasi\IEEEauthorrefmark{1}}

\IEEEauthorblockA{\IEEEauthorrefmark{1} CISPA - Helmholtz Center for Information Security, firstname.lastname@cispa.de}
\IEEEauthorblockA{\IEEEauthorrefmark{2} Cyber-Defence Campus, armasuisse Science + Technology, firstname.lastname@armasuisse.ch}
\IEEEauthorblockA{\IEEEauthorrefmark{3} University of Hamburg, firstname.lastname@uni-hamburg.de}
}

\maketitle

\begin{abstract}
Substantial reduction in launch and manufacturing costs has resulted in the accelerated deployment of small satellite missions, with commercial off-the-shelf (COTS) components becoming the prevailing standard for specific subsystems. However, this modular architecture introduces critical security risks, most notably in the Communication Subsystem (COM), which is continuously exposed by design and implicitly trusted as the entry point for command and control. We construct a tailored threat taxonomy for attacks targeting the COM subsystem and analyze representative COM systems from various vendors. Our findings uncover severe vulnerabilities across firmware, protocols, and architectural designs. This work presents the first in-depth security evaluation of widely deployed COTS COM modules employed in small satellites, identifying vulnerabilities affecting dozens of missions. To assess the real-world impact, we correlate our discoveries with open-source telemetry data, inferring at least 28 vulnerable missions in orbit that are susceptible to hostile takeover. Our work reveals that satellite COM subsystems form an attractive and dangerously neglected attack surface, necessitating urgent attention from the community.

 \end{abstract}

\IEEEpeerreviewmaketitle

\section{Introduction}
\label{sec:introduction}
The New Space era, starting in the 2000s~\cite{manulis2021cyber}, transformed small satellite missions through dramatically reduced launch and manufacturing costs. Both civil and military operators increasingly adopted commercial off-the-shelf (COTS) components and rideshare launches as bespoke satellite designs and missions became economically uncompetitive in a much-accelerated innovation environment. This commodification of, especially, \leo{} platforms has driven rapid deployment, affordable in-orbit replacements of existing satellite platforms, and a shift away from comprehensive fault handling. The result is a modular architecture in which subsystems for attitude control, energy management, communications, and data processing are purchased as flight-proven building blocks and integrated by mission-specific software.

Unfortunately, satellite systems are increasingly targeted in cyber attacks and operations~\cite{attack_on_viasat,saalman_saveljeva2023mapping,symantec_thrip_2018,flaherty2014chinese}, and New Space systems are especially vulnerable~\cite{manulis2021cyber}. While modularity accelerates development, it also relocates the security challenge to the onboard \cdhs{} software on a satellite platform. Each \cdhs{} must be customized to coordinate heterogeneous subsystems, and the absence of standardization creates different vulnerabilities across many missions~\cite{willbold2023space,scharnowski2023case}. In contrast, widely reused subsystems offer adversaries a scalable attack surface.

Among these widely-reused subsystems, the \com{} is particularly critical and at the same time the most exposed. First, the \com{} necessarily remains continuously accessible for mission control and cannot be shut off like other subsystems. Second, it provides the only uplink and downlink interface for commanding the satellite, effectively acting as the perimeter guard to an internally fully trusted satellite system. Third, as we show, this guard function is often poorly implemented, and the \com{} facilitates attacks on internal components. In combination, adversaries can therefore acquire popular \cots{} \com{} hardware, reverse-engineer its firmware and protocols, and exploit the found weaknesses on deployed commercial constellations used by a wide range of civilian and military actors~\cite{peled2023evaluating}. In addition, vulnerabilities are difficult to patch in-orbit due to data rate and safety constraints.

Even worse, \com{} vulnerabilities in satellites are not simply limited to ground-based attacks. Any other satellite equipped with a compatible transceiver may launch lateral attacks on nearby spacecraft. This satellite-to-satellite threat has received little academic attention despite its postulated feasibility~\cite{falco2020sat2sat,pavur2019cyberasat,rawlins2022death}.

While significant academic research has been conducted in recent years on the security of \cdhs{} and satellite communication itself (e.g.,~\cite{pavur2020ataleof,pavur2019cyberasat,willbold2023space,willbold2024vsaster}), the most critical onboard communications modules at the very heart of every satellite mission have not undergone direct security evaluation.

In this paper, we fill that gap. We develop a threat taxonomy tailored to spaceborne communication subsystems and select representative examples from common vendors used in over one hundred missions to perform an in-depth security assessment, mapping the findings to our taxonomy. Among other weaknesses, we discover a critical flaw that we call \textbf{\satbleed{}} in a widely-used \com{} subsystem. This vulnerability allows an unauthenticated attacker to extract cryptographic material from ground stations by simply imitating a satellite.

We correlate our findings with open-source telemetry from the \satnogs{} platform~\cite{satnogs}, a global network of over 400 ground stations, and vendor disclosures to quantify real-world impact on commercial and military missions. Using this approach, we infer that at least 28 satellites are likely vulnerable to complete, persistent takeover, with additional ones exposed to other threats discussed in this work.

Through these insights, we highlight critical vulnerabilities in small-satellite \com{} subsystems and derive architectural recommendations to improve resilience in the New Space era. Overall, we aim to strengthen the security of the small satellite ecosystem through the following contributions:

\begin{itemize}
    \item We identify a major research gap in the security of \com{} subsystems in small satellites, an area largely overlooked in academia and industry.
    \item We introduce a threat taxonomy tailored to \ac{ttnc} \com{} subsystems.
    \item We select representative \com{} modules from widely used vendors and conduct the first in-depth security evaluation of such subsystems in small satellites, revealing previously undocumented vulnerabilities.
    \item We are the first to link discovered satellite vulnerabilities to real-world deployments using open-source satellite registry data to identify affected missions.
    \item We introduce a physical-layer fuzzer, \fuzzername{}, capable of triggering bugs in sync-word detection in \coms{}\footnote{\url{https://github.com/EmsecCispa/SatBleed}}.
\end{itemize}

\section{Background}
\label{sec:background}

Small satellite missions involve spacecraft with a mass of up to 500 kg.
As of April 2025, 12,149 active satellites are in orbit, and 79\% of technology development satellites launched before 2023 weigh under 300 kg~\cite{nano2025howmanysats}. CubeSats alone make up 2,730 satellites~\cite{nanosatsFigures}. Component interoperability across small satellite form factors and the widespread reuse of identical modules amplify the attack surface shared by many missions.

\subsection{Satellite Communication}

Satellites traditionally consist of two main parts: the \emph{platform} (or satellite bus), which houses core systems such as power, attitude control, and data handling; and the  \emph{payload}, which performs the mission‑specific functions. Depending on design, platform and payload may be entirely isolated and connected only by simple control voltages or so tightly integrated that the physical boundary is indistinguishable.

Operations rely on a ground segment comprising a \mcs{} and one or more \acp{gs}. The \mcs{} schedules and dispatches \acp{tc} to trigger satellite functions, separate from any end-user data communication. \acp{gs} uplink \acp{tc} and downlink \tm{} during orbital passes, which for \leo{} last only minutes and occur several times per day. As the satellite traverses its orbit, directional antennas at both ends maintain the link; if the platform cannot successfully adapt its attitude, an omnidirectional (usually lower frequency) transceiver often serves as a backup.

In terrestrial radio installations, these \ttnc{} stations are distinct from payload communications stations. Satellite payloads employ high-data-rate transmitters or transponders that relay user traffic without further processing. Besides that, there is a \com{} for controlling the satellite itself. Payload settings may be adjusted via the dedicated payload link or the platform \com{} interface.

Satellites often enter beacon mode to maintain connectivity between passes, broadcasting spacecraft status at regular intervals. These so-called beacons are mandatory on amateur‑band missions and are used for signal acquisition by the \ac{gs}. Often beacons carry a small amount of status information about the satellite that is picked up by third‑party networks (e.g., SatNOGS~\cite{satnogs}, TinyGS~\cite{tinygs}). \ttnc{} links are typically protected by encryption or authentication mechanisms.

\subsection{\smallsat{} Subsystems}
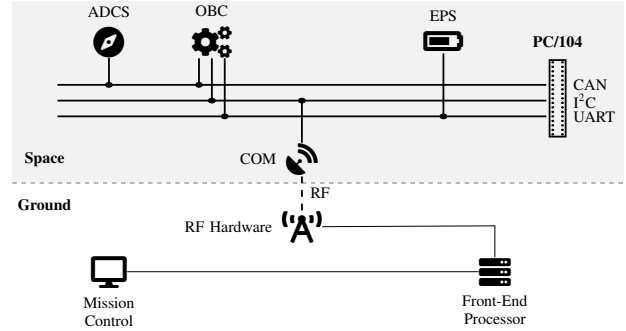
\begin{figure}[t]
  \centering
  \resizebox{0.95\columnwidth}{!}{\begin{tikzpicture}[
        >=Stealth,
        every node/.style = {font=\scriptsize, align=center},
        icon/.style       = {inner sep=0pt, font=\Large},
        busline/.style    = {thick},
      ]

      \begin{scope}[yscale=0.7]

\draw[dotted, line width=2pt, gray!60] (-4.5,0) -- (5,0);

\node at (-4, -0.5) {\textbf{Ground}};

\node[icon] (mcs) at (-3,-2) {\faDesktop};
      \node[below=0.2mm of mcs] {Mission\\Control};

\node[icon] (fep) at ( 3,-2) {\faServer};
      \node[below=0.2mm of fep] {Front-End\\Processor};

\draw[-] (mcs) -- node[below=0.3mm]{} (fep);

\node[icon] (gant) at (0,-1) {\faBroadcastTower};
      \node[left=0.2mm of gant] {RF Hardware};

\draw[-] (fep.north) -- ++(0,+0.7) -- node[above, xshift=6mm]{} (gant.east);

\fill[gray!10] (-4.5,0) rectangle (5,4);

      \node at (-4, 0.5) {\textbf{Space}};

\node[icon] (com) at (0,0.5) {\faSatelliteDish};
      \node[left=0.2mm of com] {COM};

\draw[-, dashed, thick] (gant.north) -- node[right]{RF} (com.south);

\coordinate (busL) at (-3.8,1.8);
      \coordinate (busR) at ( 3.8,1.8);

      \def\dyCAN{ 0.35}   \def\dyIIC{ 0.00}   \def\dyRS { -0.35}

\foreach \off/\lbl in {\dyCAN/CAN, \dyIIC/I\textsuperscript{2}C, \dyRS/UART}{
        \draw[busline] ($(busL)+(0,\off)$) -- ($(busR)+(0,\off)$);
        \node[right=3mm] at ($(busR)+(0,\off)$) {\lbl};
      }

\node[icon] (adcs)  at (-3,3.1) {\faCompass};
      \node[above=0.2mm of adcs] {ADCS};

      \node[icon] (obc)   at (-1.4,3.1) {\faCogs};
      \node[above=0.2mm of obc] {OBC};

      \node[icon] (power) at ( 2.2,3.1) {\faBatteryThreeQuarters};
      \node[above=0.2mm of power] {EPS};

\draw[busline] (adcs.south) -- ++(0,-0.4) -- (-3,1.8+\dyCAN);
      \filldraw[black] (-3,1.8+\dyCAN) circle (1.5pt);

\draw[busline] ($(obc.south)+(-0.2, 0)$) -- ++(-0.0,-0.4) -- (-1.6,1.8+\dyCAN);
      \filldraw[black] (-1.6,1.8+\dyCAN) circle (1.5pt); \draw[busline] (obc.south) -- ++(0,-0.4) -- (-1.4,1.8+\dyIIC);
      \filldraw[black] (-1.4,1.8+\dyIIC) circle (1.5pt); \draw[busline] ($(obc.south)+(0.2, 0)$) -- ++(0,-0.4) -- (-1.2,1.8+\dyRS);
      \filldraw[black] (-1.2,1.8+\dyRS) circle (1.5pt);

\draw[busline] (com.north) -- ++(0,+0.4) -- (0,1.8+\dyIIC);
      \filldraw[black] (0,1.8+\dyIIC) circle (1.5pt);

\draw[busline] (power.south) -- ++(0,-0.4) -- (2.2,1.8+\dyRS);
      \filldraw[black] (2.2,1.8+\dyRS) circle (1.5pt);

\foreach \i in {0,...,16} {
        \pgfmathsetmacro{\y}{0.5+\dyCAN - 0.05*\i*2}
        \filldraw[black] ($(busR)+(0.1,\y)$) circle (0.4pt);  \filldraw[black] ($(busR)+(0.25,\y)$) circle (0.4pt); }

\draw[thick] ($(busR)+(0.05, 0.9)$) rectangle ($(busR)+(0.3, -0.8)$);

\node[font=\scriptsize, above=1.4mm] at ($(busR)+(0.175, 0.8)$){\textbf{PC/104}};

\end{scope}
    \end{tikzpicture}}\caption{Ground segment (bottom) and Space segment (top).
           We illustrate the composition of subsystems in a satellite using PC/104.}
  \label{fig:pc104_bundle_split_bylink}
\end{figure}
 Unlike larger space missions that typically rely on a more monolithic spacecraft platform from a prime contractor and a payload integrated into it, smaller missions are often assembled from discrete subsystems by the future mission operator itself. Consequently, the conventional distinction between the platform and the payload becomes less pronounced. Instead, small satellites are typically constructed from a collection of interoperable components. These subsystems are selected and integrated based on the specific mission requirements. They are highly integrated and self-contained.

Figure~\ref{fig:pc104_bundle_split_bylink} illustrates the space/ground segment architecture of a small satellite mission.
Commonly, subsystems are housed in cuboid enclosures with two PC/104 connectors for vertical stacking. The remaining three sides are often used for auxiliary wiring, such as coaxial cables between modules or point-to-point data lines. These interfaces are typically vendor-specific and not interoperable, making the PC/104 connector the primary communication interface within the stack. While PC/104 provides a mechanical and electrical framework, the actual pin usage is not strictly standardized. Nevertheless, interface conventions exist, with shared buses such as I\textsuperscript{2}C, CAN, and UART (e.g., RS-485) and power rails appearing in predictable locations.

Thus, integrators typically cannot choose which pins are used for which interfaces, leading to multiple devices sharing a bus. This results in a lack of interface isolation, increasing the risk of compromised subsystems, allowing attackers to laterally move to other subsystems.

\subsubsection{Subsystem Taxonomy}
We follow the taxonomy by Willbold et al.~\cite{willbold2023space} and focus on the security-relevant subsystems:\\
\paragraph{\textbf{\acf{obc}}}
    \acp{obc} provide general-purpose computing capabilities to satellites. There are always-on \acp{obc}, which run critical command handlers and monitor the satellite platform. Typically, this type of \obc{} utilizes microcontrollers with embedded real-time operating systems such as FreeRTOS~\cite{amazon2023freertos}, Zephyr~\cite{zephyrRtos}, or RTEMS~\cite{rtems}.
    Additionally, there are high-end \acp{obc} that may complement the aforementioned systems. These units run on more compute-optimized microprocessors, often Arm Cortex-A SoCs~\cite{AlenSpaceTrevo}. Frequently, these processors are either part of the payloads or integrated with high data rate transceivers in conjunction with FPGA fabric.\\
\paragraph{\textbf{\acf{com}}}
    The \com{} is the gateway between the outside world and the rest of the satellite platform. Thus, it is a basic requirement for the satellite to operate. Usually, small satellites have both omnidirectional communications on UHF/VHF and, optionally, if higher data rates are required, unidirectional antennas with transceivers working in S/X/Ku/Ka-Band. Notably, the same subsystems are often used for the ground-station part of the communication link under the name of \fep{}.

    Software and Configuration updates to the \com{} subsystem in particular are risky, as a faulty patch can sever the \ttnc{} link and incapacitate fundamental operations terminally, e.g., thermal regulation, or attitude/orbit control. At the same time, other research has shown that efficient over-the-air (OTA) update mechanisms in general remain an open challenge in these constrained environments~\cite{rederlechner2026one}.\\
\paragraph{\textbf{\acf{cdhs}}}
    The \cdhs{} subsystem is the combination of hardware and software that processes telecommands. This typically includes software running on an always-on low-power \obc{} as well as hardware on the satellite that handles low-level commands, particularly for scenarios like hardware failover. In this paper, when referring to the \cdhs{}, we are discussing the system that implements operational commands and executes logic for station-keeping tasks required by the satellite. For \cdhs{} software, there are several open-source and proprietary suites available~\cite{mccomas2016core, Bocchino2018FPrime, raccoonos}. Previous papers have already studied the security of those systems and have concluded that there are general security concerns associated with them~\cite{scharnowski2023case, willbold2023space}. For malicious inputs to reach this hardware and software, they previously have to pass through the COM. This means that in case the \com{} is enforcing authentication, unauthenticated attackers do not have the capability to reach these components.\\
\paragraph{\textbf{\acf{adcs}}}
    The \adcs{} dictates the attitude of the satellite, determining the direction in which it is pointing. To achieve this, it uses sensors to read the current attitude, utilizing the Earth's magnetic field and sun sensors for rough pointing. For more precise positioning, it may employ star trackers.
    Control over the attitude is then performed by magnetorquers (electromagnets that react against the Earth's magnetic field). These are either used to directly influence the attitude or to desaturate reaction wheels, which can make changes to the attitude more quickly~\cite{magnetorquer}.
    This makes the \adcs{} an interesting target for attackers, as they need control over a satellite's attitude to point a directional antenna, prevent sensors from capturing data, or deliberately cause damage to the spacecraft.\\
\paragraph{\textbf{\acf{eps}}}
    The \eps{} is the system that coordinates the generation and use of electrical power. This includes control over solar charging and switching power rails. Usually, the EPS also handles keeping the batteries within their thermal limits, activating electric heaters for the battery cells when necessary. This makes it an interesting target for sabotage if an attacker aims to artificially cause a satellite mission to fail prematurely.

\subsection{Satellite Network Architecture}

The simplest type of frame in a space protocol encapsulates a single packet containing one command. This model abstracts the satellite as a command-driven device, receiving TCs as the uplink from the ground segment and providing a response via TM as the downlink \cite{esa_tmtc}.

Internally, these commands are disseminated through the onboard communication architecture and parsed by the subsystems responsible for their execution or directly interpreted by the COM. A common drawback of satellite networks is the lack of privilege separation within the onboard software stack. All commands are treated with equal authority once accepted by the satellite, lacking separation of different privilege levels to limit access to sensitive commands and telemetry data.
We discuss some specific space protocol designs in Section~\ref{sec:case_study}.

\section{Criticality of the \com{} System for Satellites}
\label{sec:criticality}
We discuss the motivation for examining the security of \com{} systems, focusing on their critical position within the wider satellite ecosystem and the existing research gaps surrounding them.

\subsection{Systematization of Satellite Security Research}

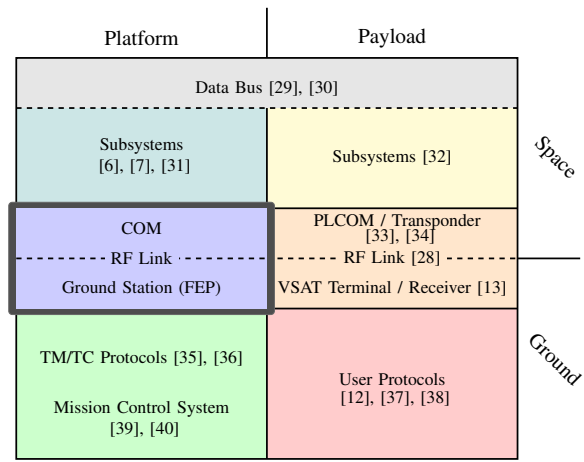
\begin{figure}[t!]
    \centering
    \resizebox{0.9\columnwidth}{!}{\begin{tikzpicture}
\tikzstyle{every node}=[font=\normalsize]
            \begin{scope}[yscale=0.8]

\fill[gray!20]  (3.75,19.5)   rectangle (13.75,18.25);
\fill[teal!20]  (3.75,18.25)  rectangle (8.75,15.75);
\fill[yellow!20](8.75,18.25)  rectangle (13.75,15.75);

\fill[red!20]   (8.75,13.25)  rectangle (13.75,9.5);    \fill[orange!20](8.75,15.75)  rectangle (13.75,13.25);  \fill[green!20] (3.75,13.25)  rectangle (8.75,9.5);

\draw [line width=1pt] (3.75,19.5) rectangle (13.75,9.5);
            \draw [line width=1pt, -, >=Stealth] (13.75,14.5) -- (15,14.5);
            \draw [line width=1pt, -, >=Stealth] (8.75,19.5) -- (8.75,20.75);

\node [font=\large] at (6.25,20)  {Platform};
            \node [font=\large] at (11.25,20) {Payload};

\draw [line width=1pt, dashed] (3.75,18.25) -- (13.75,18.25);

\draw [line width=1pt] (8.75,18.25) -- (8.75,9.5);

\draw [line width=1pt, dashed] (8.75,14.5) -- (13.75,14.5) node[pos=0.5, fill=orange!20]{RF Link \cite{tedeschi2022satellite}};

\draw [line width=1pt] (3.75,15.75) -- (13.75,15.75);
            \draw [line width=1pt] (3.75,13.25) -- (13.75,13.25);

            \draw[draw=black!60, fill=black!70, rounded corners=2pt] (3.6,15.9) rectangle (8.9,13.1); \fill[blue!20]  (3.75,15.75)  rectangle (8.75,13.25);   \draw [line width=1pt, dashed] (3.75,14.5) -- (8.75,14.5) node[pos=0.5, fill=blue!20]{RF Link}; \node at (8.75,18.75) {Data Bus~\cite{driouch2024distids,he2019security}};

\node at (6.25,17.30) {Subsystems};
            \node at (6.25,16.80) {~\cite{willbold2023space,scharnowski2023case,jero2024securing}};
            \node at (11.25,17){Subsystems~\cite{deadf00dHackedESAs}};

\node at (6.25,15.25) {COM};
            \node at (11.35,15.4){PLCOM / Transponder};
            \node at (11.35,15.05){~\cite{smailes2023waththis,wigchert2025detection}};

\node at (6.25,13.75) {Ground Station (FEP)};
            \node at (11.25,13.75){VSAT Terminal / Receiver~\cite{willbold2024vsaster}};

\node at (6.25,12) {TM/TC Protocols~\cite{masson2024sdlsimpl,willbold2024satellite}};
            \node at (11.25,11.50){User Protocols};
            \node at (11.25,11.0){\cite{pavur2020ataleof,pavur2019secrets,Pavur2021QPEPAnActionable}};

\node at (6.25,10.75) {Mission Control System};
            \node at (6.25,10.25){\cite{visionspaceYamcsVulns,honeysat}};

\node [font=\large, rotate around={-45:(0,0)}] at (14.5,17) {Space};
            \node [font=\large, rotate around={-45:(0,0)}] at (14.5,12) {Ground};

        \end{scope}
        \end{tikzpicture}}

    \caption{Illustration of the different aspects of space cyber security research. Different targets of space cyber security research have been colored. The highlighted area represents a research gap we aim to fill.}
    \label{fig:space_cyber_fig}
\end{figure}

 First, we can distinguish \textit{platform} and \textit{payload} with their respective subsystems. Second, we can distinguish ground and space systems. Finally, data buses, interfaces, and links between space and ground are a separate concern. Based on these dimensions, we create a taxonomy comprising all aspects of these subsystems and systematize existing research as illustrated in  Figure~\ref{fig:space_cyber_fig}.

\subsubsection{Space}
\begin{description}
\item[Data Bus:] Existing work has analyzed vulnerabilities of high-trust data buses such as CAN or MIL-STD-1553B used in Satellites \cite{driouch2024distids,he2019security}. Similar to unauthenticated use of such bus systems in aircraft or cars, an attacker with access to the bus can conduct \ac{dos} and spoofing attacks.
\item[Platform: Subsystems] Prior studies systematically analyzed onboard platform components and flight software such as On-Board Computers, Attitude Determination and Control Systems, and Electrical Power Systems, revealing vulnerabilities in firmware update mechanisms and insufficient interface isolation~\cite{willbold2023space,scharnowski2023case,jero2024securing}.
\item[Payload: Subsystems] Previous research on space-segment payload components, including high-data-rate transceivers and sensor interfaces, has identified implementation flaws, and parsing errors~\cite{deadf00dHackedESAs}.
\end{description}
\subsubsection{Ground}
\begin{description}
\item[Platform: TM/TC Protocols] Link-layer Telecommand and Telemetry protocols have been subjected to replay, spoofing, and framing attack analyses, which demonstrate the prevalence of weaknesses in authentication and integrity checks \cite{masson2024sdlsimpl}, although some implementations have shown fewer obvious weaknesses \cite{willbold2024satellite}.

\item[Platform: Mission Control Systems] Security evaluations of Mission Control System software have uncovered insufficient access control mechanisms, vulnerabilities in scheduling components, and inadequate logging practices, all of which enable potential unauthorized command injection \cite{visionspaceYamcsVulns}. Other works have captured malicious interactions with exposed MCS infrastructure~\cite{honeysat}.

\item[Payload: User Protocols] Investigations into end-user satellite broadband and data-service protocols have demonstrated \ac{dos}, and confidentiality flaws, leading to proposals for protocol hardening and enhanced encryption strategies \cite{pavur2020ataleof,pavur2019secrets,Pavur2021QPEPAnActionable}.
\end{description}

\subsubsection{Space-Ground Interfaces}
\begin{description}
\item[RF Links:] Researchers have investigated payload RF links~\cite{salkield2023firefly}. Wider surveys on SATCOM RF security such as \cite{tedeschi2022satellite} discuss the issue more generally from a lower-layer perspective.

\item[Payload: PLCOM / Transponder] Detailed analyses of payload transponders have revealed that their radio frequency links are susceptible to link-layer attacks such as frame manipulation and jamming \cite{smailes2023waththis,wigchert2025detection}.

\item[Payload: VSAT Terminal / Receiver] Assessments of Very Small Aperture Terminal (VSAT) receivers have exposed hardware and firmware vulnerabilities that adversaries could exploit to compromise commercial user terminals~\cite{willbold2024vsaster,wireless2024bisping}.

\item[Platform: COM / \ttnc{} Ground Station] Currently, there is no security evaluation of TT\&C transceivers or ground-station RF hardware in the literature, as we discuss next.
\end{description}

\subsection{Research Gap: \com{} Systems and \ttnc{} Links}

As shown in Figure~\ref{fig:space_cyber_fig} and our systematization, existing work covers each quadrant, yet we observe a distinct gap in the platform's space-ground interface, even though it represents a high-impact attack surface. There is no work investigating COM components, either space-based or ground-based (\fep{}), used for TT\&C, even though they are the ideal targets for attackers to conduct malicious actions. Possible reasons for this might be that in larger customized satellites, these subsystems are highly sensitive, and therefore, gaining information on them is difficult.

\subsection{Importance of the Communication System}

Among all subsystems, the \com{} is paramount, as it provides the sole link for command uplink, telemetry downlink, and, in some cases, internal message routing between modules. It thus occupies a central position guarding the satellite’s perimeter. The \com{} is inherently exposed via an external RF link and therefore cannot be shielded by additional filtering, while maintaining direct access to internal data buses where subsystems exchange messages under the assumption of a trusted environment.

In practice, attitude-control limitations and imprecise antenna pointing lead many satellites to rely on omnidirectional links~\cite{saeidi2025enhancing}. While these allow signal reception without active pointing, they also enable drive-by exploitation. When an adversary compromises the \com{} subsystem, they can control telemetry and telecommand exchange with the ground segment. Even a simple \ac{dos} vulnerability may render a satellite uncontrollable and cause mission failure.

The costly, slow, and one-shot nature of launches, combined with reliability requirements, often leads engineers to prioritize safety over security. As a result, diagnostic backdoors and "clear modes"~\cite{aac_pulsar_advertisement_card,sdls} are routinely included to ensure mission continuity. Regarding link protection, little research evaluates the correctness of these often proprietary protocols.

\com{} subsystems vary significantly in complexity. Simple transceivers send raw frames in near real time, while more advanced units parse packets, handle fragmentation, and maintain routing tables. Store-and-forward devices buffer telemetry for scheduled downlink. These systems may be the only barrier preventing unauthenticated control of the satellite. If such functionality is not hardened, the attack surface grows substantially.

Similar concerns apply to RF transceivers in commercial \gstation{} appliances, often called \fep{}. Vendors typically provide matching \gstation{} hardware for their \com{} subsystems, connected to a RF front-end to communicate with the satellite. This tight coupling means both the spacecraft and its \ttnc{} \gstation{} may share vulnerabilities. Attacks from ground-based adversaries targeting the \gstation{} and space-based adversaries targeting the spacecraft can remain stealthy, as they can use very low RF power, often undetectable by monitoring authorities.

A final concern is the opacity of the software executing on \coms{}. Once a component of a satellite has been compromised, it is almost impossible to reset because the usual recovery mechanisms are built with safety in mind and do not withstand an adversary who deliberately manipulates a system. With subsystems allowing lateral movement, an attacker could still keep persistent access without further measures. None of the \coms{} analyzed features a resilient fallback mechanism.

This paper therefore presents a detailed analysis of hardware, communication protocols, firmware, and operational practices.

\section{\com{} Subsystem Threats}
\label{sec:threatmodel}

As attacks on space-based assets evolve, a new attacker ecosystem may emerge. The increasing number of spacecraft, and thus identical \coms{}, enables attack scenarios that depend on a high density of vulnerable satellites.

In this section, we argue that the \com{} is central to these emerging threats and outline how a more mature attack ecosystem could develop as adversaries pivot to vulnerable small satellites. Given the typically limited lifetime of LEO small-satellite missions, these risks can still be addressed in future designs before widespread exploitation occurs.

\subsection{Novel Threats to Space-Based Assets}
\label{sec:novel_threats}
\subsubsection{Space Botnets}
A monoculture of \coms{} implies that a single vulnerability can rapidly compromise a substantial fraction of the satellite population. The threat is exacerbated for platforms that are able, either by design or by coincidence, to conduct satellite‑to‑satellite communications. Under such conditions, self‑propagating "space worms" can emerge, exploiting the same flaw moving between satellites.
Propagation speed is only limited by orbital dynamics and the number of targets. Once initiated, the process is effectively unstoppable and may eventually encompass the complete satellite population, potentially without immediate observable effects. The outcome is a large‑scale, practically untraceable attack against space‑based infrastructure.
Similar to DoS botnets on the internet, constellations of radio-frequency jammers consisting of compromised \com{} systems could show up in space, as suggested in~\cite{rawlins2022death}. Once activated, these devices could emit interference that cannot be disabled from the ground. Such a network could systematically block critical downlink or uplink signals~\cite{planta2024let}, thereby disrupting mission operations or degrading the service quality of other satellites. The effectiveness of this threat is amplified, as the \com{} subsystem usually does not tolerate any other transmitter on the same channel as stated in Section~\ref{subsec:gom}.

\subsubsection{Satellite Spyware} Satellites are designed with the expectation that subsystems may fail, making redundancy a key architectural principle. To enable rapid fault recovery, platforms include multiplexers that allow operators to switch from a primary component to a cold-redundant backup. An advanced adversary can exploit this mechanism to establish persistence. After compromising a \com{} or \obc{}, malware may laterally move to a normally powered off backup unit. The attacker can then simulate a failure of the primary element, enabling persistent use of the satellite’s main functionality without operator awareness.

The attacker can hide additional power consumption by impersonating the \eps{} on spacecraft data buses or by corrupting \eps{} telemetry so reduced battery capacity appears as natural degradation. From the backup unit, spyware can passively collect payload data, inject falsified commands, and exfiltrate information via covert channels embedded in normal downlink frames. Detection is difficult because the redundant unit is expected to remain offline unless needed and may appear to have simply failed. As a result, operators may never realize the attacker maintains a foothold that can be reactivated at any time.

\subsection{\com{} Subsystem Sub-Components}

\begin{figure}[t]
  \centering
  \resizebox{1.0\columnwidth}{!}{
    \begin{tikzpicture}[
  font=\LARGE,
  box/.style={draw, rounded corners=2pt, minimum width=4.4cm, minimum height=1.9cm, align=center, line width=3pt},
  colsep/.store in=\colsep, colsep=1.8cm,   rowsep/.store in=\rowsep, rowsep=3.0cm    ]

\definecolor{SigAColor}{HTML}{1F77B4} \definecolor{SigBColor}{HTML}{FF7F0E} \definecolor{SigCColor}{HTML}{2CA02C}

\tikzset{
  siguauth/.style={->, draw=SigAColor, line width=3.0pt, >={Triangle[length=3mm]}},             sigauth/.style={->, draw=SigBColor, line width=3.0pt, dashed, >={Triangle[length=3mm]}},sigconf/.style={->, draw=SigCColor, line width=3.0pt, dotted, >={Triangle[length=3mm]}}      }

\coordinate (frameNW) at (-9.0,2.0);
\coordinate (frameSE) at (9,-6.5);
\draw[draw=black!60, thick, fill=black!10, rounded corners=2pt] (frameNW) rectangle (frameSE);
\node[box] (intcmd) {Internal\\Commands};
\node[box, left=\colsep of intcmd] (rfproc) {Frame Processing};
\node[box, right=\colsep of intcmd] (dbg) {Debug / Config\\Interface};

\node[box, below=\rowsep of rfproc] (auth) {Authentication};
\node[box, xshift= -20pt, below=0.5cm of dbg] (intauth) {Authentication};
\node[box, below=\rowsep of intcmd] (route) {Routing};
\node[box, below=\rowsep of dbg] (netif) {Network\\Interfaces};

\node[box, above=2.5cm of rfproc, xshift= -30pt] (bb) {Baseband\\Processing};
\node[box, right=0.8cm of bb] (rf) {RF-Frontend};

\coordinate (frameSW) at (frameNW |- frameSE);
\coordinate (frameNE) at (frameSE |- frameNW);
\node[anchor=north east] at ([xshift=-2pt,yshift=-2pt]frameNE) {\textbf{Software}};
\node[draw, minimum width=0.55cm, minimum height=0.45cm, inner sep=0pt,
      right=0.8cm of dbg] (pTop) {\LARGE\attackerPA{}};
\node[draw, minimum width=0.55cm, minimum height=0.45cm, inner sep=0pt,
      right=0.8cm of netif] (pBot) {\LARGE\attackerMI{}};

\node[box, right=5.0cm of rf] (com) {FEP};

\node[draw, minimum width=0.55cm, minimum height=0.45cm, inner sep=0pt,
      right=0.8cm of com] (pFep) {\LARGE\attackerMI{}};

\node[draw, minimum width=0.55cm, minimum height=0.45cm, inner sep=0pt, yshift=-38pt,
      right=0.8cm of rf] (pCom) {\LARGE\attackerMT{}};

\node[draw, minimum width=0.55cm, minimum height=0.45cm, inner sep=0pt, yshift=38pt,
      left=0.8cm of com] (pRF) {\LARGE\attackerGSA{}};

\node[draw, minimum width=0.55cm, minimum height=0.45cm, inner sep=0pt,
      left=1.7cm of com, yshift=10pt] (pMiddle) {\LARGE\attackerAitM{}};

\draw[siguauth, transform canvas={yshift= 15pt}] (rf.west) -- (bb.east);
\draw[sigauth, transform canvas={yshift= -15pt}] (bb.east) -- (rf.west);
\draw[sigconf] (bb.east) -- (rf.west);

\draw[sigconf] (rfproc.east) -- (rf.south);
\draw[siguauth, transform canvas={xshift= 10pt}] (bb.south) -- (rfproc.north);

\draw[sigconf, transform canvas={xshift= -10pt}] (rfproc.north) -- (bb.south);
\draw[sigauth, transform canvas={xshift= -30pt}] (rfproc.north) -- (bb.south);

\draw[siguauth] (pTop.west) -- (dbg.east);

\draw[siguauth, transform canvas={xshift= 10pt}] (rfproc.south) -- (auth.north);
\draw[sigauth, transform canvas={xshift= -10pt}] (auth.north) -- (rfproc.south);

\draw[sigauth, transform canvas={yshift= 10pt}] (auth.east) -- (route.west);
\draw[siguauth, transform canvas={yshift= -10pt}] (route.west) -- (auth.east);

\draw[siguauth] (route.north east) -- (intauth.south west);
\draw[sigauth] (intauth.north west) -- (intcmd.south east);

\draw[<->, draw=SigAColor, line width=3pt, >={Triangle[length=3mm]}] (route.east) -- (netif.west);

\draw[siguauth, transform canvas={yshift= 10pt}] (dbg.west) -- (intcmd.east);
\draw[sigconf, transform canvas={yshift= -10pt}] (intcmd.east) -- (dbg.west);

\draw[siguauth] (pBot.west) -- (netif.east);

\draw[siguauth, line width=1mm] (pRF.north east) -- (com.north west);
\draw[sigauth, line width=1mm] (com.north west) -- (pRF.south east);
\draw[siguauth, line width=1mm] (pCom.west) -- (rf.south east);

\draw[sigauth, line width=1mm, transform canvas={yshift= -10pt}] (rf.east) -- (com.west);

\draw[sigauth, line width=1mm, transform canvas={yshift= -20pt}] (com.west) -- (rf.east);

\draw[sigauth] (com.west) -- (pMiddle.north east);
\draw[siguauth] (pMiddle.south east) -- (com.west);

\draw[sigauth] (rf.east) -- (pMiddle.south west);
\draw[siguauth] (pMiddle.north west) -- (rf.east);

\draw[siguauth] (pFep.west) -- (com.east);

\draw[sigconf] (intcmd.south) -- (route.north);

\draw[sigconf] (intcmd.south) -- (intauth.west);

\draw[sigconf] (intcmd.west) -- (rfproc.east);

\draw[sigconf] (intcmd.south west) -- (auth.north east);

\draw[sigconf] (intcmd.south) -- (netif.north west);

\def\L{55mm}  \def\S{6mm}

\coordinate (legendOrigin) at ($ (frameSW)!0.5!(frameSE) + (-90mm,-12mm) $);

\coordinate (LA) at (legendOrigin);
\coordinate (LB) at ($(legendOrigin)+(\L+\S,0)$);
\coordinate (LC) at ($(legendOrigin)+(2*\L+2*\S,0)$);

\draw[siguauth] (LA) -- ++(\L,0) node[midway,above=1mm]{Unauthenticated Data};
\draw[sigauth] (LB) -- ++(\L,0) node[midway,above=1mm]{Authenticated Data};
\draw[sigconf] (LC) -- ++(\L,0) node[midway,above=1mm]{Internal Configuration};

\end{tikzpicture}

  }
  \caption{Sub-component based view of data/attack flow for the Communications Subsystem with secure reference data flows.}
  \label{fig:taxonomy}
\end{figure}
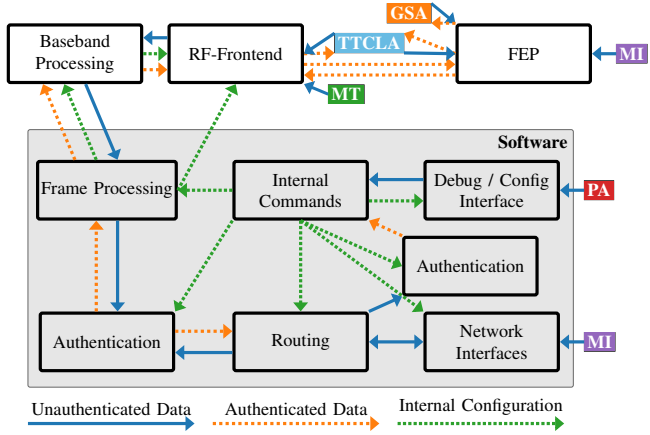
 To better understand architectural weaknesses in \com{} systems and enable a systematic analysis, we derive a threat taxonomy. Existing taxonomies, such as \cite{willbold2023space}, operate at the component level and lack the granularity needed to capture all threats to the \com{} and corresponding \fep{}. To address this, we analyze the system features of \numcomsintaxonomy{} \coms{} and derive a sub-component-based model, shown in Figure~\ref{fig:taxonomy}.

We model information flow within an idealized \com{} architecture, which can then be compared to real implementations to reveal architectural weaknesses and the consequences of vulnerabilities. Vulnerabilities can be classified by affected sub-components, and exploit chains visualized through data flow. Because \coms{} involve complex hardware-software interactions, both aspects and their interplay must be modeled.

As shown in Figure~\ref{fig:taxonomy}, numerous entry points exist for unauthenticated, potentially malicious inputs into \com{} sub-components.

\coms{} and \fep{}s face multiple threat vectors. Attackers may directly target systems via the RF link to disrupt operations or gain a foothold. Once access is obtained, they may seek persistence by embedding malicious firmware. Another risk arises when adversaries intercept and modify \tc{} or \tm{}. Based on these scenarios, classify relevant attacker types for \com{} security.

\subsection{Attacker Model}
\label{sec:attackers}
For a satellite mission that contains a model of \com{} and corresponding \fep{}, we distinguish the following attacker types according to their capabilities.

\paragraph{\attackerPA{~}~\textbf{Short Duration Physical Access Attacker}}
The attacker can tamper with hardware for a short time period before launch or during development at the supplier. This includes physically manipulating components to access interfaces that are typically not exposed.

\paragraph{\attackerAitM{}~\textbf{TT\&C Link Adversary}}
The attacker can intercept the legitimate TT\&C uplink from the ground station and operate their own ground station that can communicate with the satellite. An eavesdropper positioned outside the secured perimeter of a ground station can exploit side lobes to capture telecommands~\cite{wireless2024bisping}, analyze them, and replay them at a different time and location. Note that this attacker does not require the capability to prevent the delivery of messages transmitted.

\paragraph{\attackerMT{~}~\textbf{Malicious Transmitter}}
The attacker can transmit arbitrary messages to the target satellite or \gstation{}. Such capabilities can be obtained by repurposing amateur‑radio equipment and commercially available hardware that may be ordered online~\cite{polysat_earthstation}. This category also covers benign ground stations that are compromised and subsequently abused, as well as ground‑station‑as‑a‑service offerings operated with malicious intent. Alternatively, the attacker can operate or compromise another satellite to pose a threat via inter-satellite communication.

\paragraph{\attackerGSA{}~\textbf{Ground Station Attacker}}
The attacker can send arbitrary signals to the \gstation{} and receive responses from the \gstation{}. The GSA represents a weaker version of the {\scriptsize\attackerAitM{}} that is explicitly attacking the \gstation{} from the ground and cannot reach the satellite.

\paragraph{\attackerMI{~}~\textbf{Low Privileged Malicious Insider}}
A low‑privileged participant in a satellite mission who can cause arbitrary, validly authenticated data to be transmitted via the \com{}. This attacker might have the ability to inject data through either the satellite or the \ac{gs} communication systems. For example, they may have access to the \fep{} or to a computer running mission‑control software. Alternatively, in a multi‑payload mission, the attacker could be embedded as one of the payloads. In such cases, shared bus architectures may allow them partial access to other payloads.\\

We generally always name the least capable attacker who could perform an attack. Attackers can be ordered according to the hierarchy below, where stronger attackers imply weaker ones.\\

\begin{center}
\forestset{
  rightgrow/.style={
    for tree={
      grow=0,                    parent anchor=east,
      child anchor=west,
      anchor=west,
      edge={draw=none},          edge label={node[midway,sloped,inner sep=0pt]{$\Rightarrow$}}, l sep=16pt,                s sep=1mm,                 inner sep=1pt, outer sep=0pt,
      align=left,
    }
  }
}
\begin{forest} rightgrow
  [\attackerMI{}
    [\attackerAitM{}
      [\attackerMT]
      [\attackerGSA{}]
    ]
  ]
\end{forest}
\end{center}

\subsubsection{Modeling Restrictions}

Executing these attacks in practice requires adversaries to overcome significant operational constraints. An attacker must account for narrow timing windows, as \leo{} orbital passes typically last only minutes, depending on the orbit and the attacker's location, and occur only a few times per day. Furthermore, exercising \textbf{Malicious Transmitter}~{\scriptsize\attackerMT{}} capabilities (as the weakest attacker of this type) requires constructing a directional antenna, managing Doppler shift, and possessing sufficient transmit power to reach the satellite. While this is easier in VHF/UHF than in other bands due to lower equipment requirements and a smaller absolute Doppler shift, tuning a setup for this is not trivial. Finally, persistent malicious transmissions risk detection of their geolocation by spectrum monitoring authorities. While we abstract these factors to evaluate the underlying hardware weaknesses, they represent practical hurdles that dictate the adversary's necessary sophistication and resources.

\subsection{Generalized \com{} Threat Taxonomy}

\begin{figure*}[t]
\centering
\scriptsize
\setlength{\tabcolsep}{3pt}
\renewcommand{\arraystretch}{0.95}
\newcommand{\ccell}[1]{\makecell[c]{\renewcommand{\arraystretch}{0.9}#1}}
\setlength{\aboverulesep}{0pt}
\setlength{\belowrulesep}{0pt}
\begin{tabular}{
    >{\centering\arraybackslash}p{0.11\textwidth}
    >{\centering\arraybackslash}p{0.11\textwidth}
    >{\centering\arraybackslash}p{0.11\textwidth}
    >{\centering\arraybackslash}p{0.11\textwidth}
    >{\centering\arraybackslash}p{0.12\textwidth}
    >{\centering\arraybackslash}p{0.11\textwidth}
    >{\centering\arraybackslash}p{0.11\textwidth}
    >{\centering\arraybackslash}p{0.14\textwidth}
}
    \multicolumn{2}{c}{\textbf{Hardware}} & \multicolumn{6}{c}{\textbf{Software}} \\
    \cmidrule(r){1-2} \cmidrule(l){3-8}

    \ccell{\textbf{Baseband}\\\textbf{Processing}} &
    \ccell{\textbf{RF-Frontend}} &
    \ccell{\textbf{Internal}\\\textbf{Commands}} &
    \ccell{\textbf{Network}\\\textbf{Interfaces}} &
    \textbf{Authentication} &
    \textbf{Routing} &
    \ccell{\textbf{Frame}\\\textbf{Processing}} &
    \ccell{\textbf{Debug /}\\\textbf{Config Interface}} \\

    \cmidrule(r){1-1} \cmidrule(r){2-2} \cmidrule(r){3-3} \cmidrule(r){4-4}
    \cmidrule(r){5-5} \cmidrule(r){6-6} \cmidrule(r){7-7} \cmidrule(r){8-8}

    \ccell{\textbf{State}\\\textbf{Corruption}\\{\tiny\attackerMT{}}} &
    \ccell{\textbf{Cause Physical}\\\textbf{Damage}\\{\tiny\attackerMI{}}} &
    \ccell{\textbf{State}\\\textbf{Corruption}\foundVuln\\{\tiny\attackerMI{}\attackerPA{}}} &
    \ccell{\textbf{Packet}\\\textbf{Smuggling}\foundVuln\\{\tiny\attackerMI{}}} &
    \ccell{\textbf{Algorithm Side}\\\textbf{Channels}\foundVuln\\{\tiny\attackerAitM{}}} &
    \ccell{\textbf{Memory}\\\textbf{Corruption}\\{\tiny\attackerMI{}}} &
    \ccell{\textbf{Memory}\\\textbf{Corruption}\\{\tiny\attackerMT{}}} &
    \ccell{\textbf{Reprogramming}\foundVuln\\\attackerPA{}} \\

    \ccell{\textbf{Jamming}\\{\tiny\attackerMT{}}} &
    \ccell{\textbf{Jamming}\\{\tiny\attackerMT{}}} &
    \ccell{\textbf{Privilege}\\\textbf{Escalation}\foundVuln\\{\tiny\attackerMI{}\attackerPA{}}} &
    \ccell{\textbf{Identity}\\\textbf{Spoofing}\\{\tiny\attackerMI{}}} &
    \ccell{\textbf{Weak}\\\textbf{Authentication}\foundVuln\\{\tiny\attackerMT{}}} &
    \ccell{\textbf{Identity}\\\textbf{Spoofing}\\{\tiny\attackerMI{}}} &
    \ccell{\textbf{State}\\\textbf{Corruption}\foundVuln\\{\tiny\attackerMT{}}} &
    \ccell{\textbf{Config}\\\textbf{Modification}\\{\tiny\attackerPA{}}} \\

    \ccell{\textbf{Resource}\\\textbf{Exhaustion}\\{\tiny\attackerMT{}}} &
    \ccell{\textbf{Signal}\\\textbf{Leakage}\foundVuln\\{\tiny\attackerMT{}}} &
    \ccell{\textbf{Memory}\\\textbf{Corruption}\foundVuln\\{\tiny\attackerMI{}\attackerPA{}}} &
    \ccell{\textbf{Information}\\\textbf{Leak}\\{\tiny\attackerMI{}}} &
    \ccell{\textbf{Credential}\\\textbf{Compromise}\\{\tiny\attackerGSA{}\attackerMT{}\attackerPA{}}} &
    \ccell{\textbf{Information}\\\textbf{Leakage}\foundVuln\\{\tiny\attackerMI{}\attackerPA{}}} &
    \ccell{\textbf{Frame}\\\textbf{Smuggling}\\{\tiny\attackerMT{}}} &
    \ccell{\textbf{Memory}\\\textbf{Corruption}\foundVuln\\{\tiny\attackerPA{}}} \\

    &
    &
    &
    \ccell{\textbf{Memory}\\\textbf{Corruption}\\{\tiny\attackerMI{}}} &
    \ccell{\textbf{Memory}\\\textbf{Corruption}\\{\tiny\attackerMT{}\attackerPA{}}} &
    &
    \ccell{\textbf{Resource}\\\textbf{Exhaustion}\\{\tiny\attackerMT{}}} &
    \\
\end{tabular}
\caption{Taxonomy of subcomponents, threats, and discovered vulnerabilities. Each hypothesized \textbf{threat} (body columns) is annotated with the weakest required attacker model. Each threat with a corresponding vulnerability in Table \ref{tab:vulnerabilities} is marked with \foundVuln{}.}
\label{fig:vuln_tree}
\end{figure*}

A taxonomy of threats is visualized in Figure~\ref{fig:vuln_tree}. Building on the previously discussed novel threat scenarios, we now shift the focus from large-scale, persistent threat scenarios in Section~\ref{sec:novel_threats} to the underlying threats to the \com{} that enable such attacks.

To systematically capture these weaknesses, we derive a generalized vulnerability taxonomy for \com{} systems. The taxonomy is based on the sub-component oriented view of the \com{} introduced in Figure~\ref{fig:taxonomy}, which models both data flow and trust boundaries (e.g. via Authentication) across hardware and software elements. Each hypothesized threat in the nodes of Figure~\ref{fig:vuln_tree} is associated with the specific sub-component it affects and is further annotated with the weakest attacker model presented in Section~\ref{sec:attackers} required to exploit it. Attackers are chosen under the assumption that the underlying data flow of a \com{} system follows Figure~\ref{fig:taxonomy} and thus might allow to be exploitable by a weaker version of the annotated attackers for non-ideal implementations in real-world \com{} devices.

\section{Evaluation of COTS COMs}
\label{sec:case_study}
To evaluate the practical validity of our threat taxonomy, we apply it to three real-world \com{} subsystems. The complete data flows of the analyzed systems is presented in the appendix \ref{appendix:taxonomy}.

The threat taxonomy outlined in Figure \ref{fig:vuln_tree} actively guided the evaluation methodology in this Section. By systematically evaluating the data flows against the threat vectors defined for each sub-component (e.g., Frame Processing, Routing), we proactively identified attack classes, such as the Syncword-based DoS and architectural routing leaks, that might have been overlooked in an unstructured review. Furthermore, because this taxonomy models the core architectural elements of spaceborne communications, it serves as a generalized framework that can be applied to systematically evaluate the security of other COM subsystems across the broader NewSpace ecosystem.

We perform the evaluation under the assumption that the operators made a best effort to secure their systems; we exclude issues arising from fundamentally insecure configurations.

\subsection{Identifying Critical Communication System Vendors} \label{sec:which_coms_are_critical}

To focus our security analysis on the \com{} modules whose compromise would have the greatest operational and safety impact and to ensure our findings generalize across the small satellite market, we first define selection criteria. Using these criteria, we identify three vendors whose \com{}, providing a meaningful sample for uncovering systemic weaknesses.

\subsection*{Selection Criteria}
\label{sec:com_criteria}

\begin{itemize}
  \item \textbf{Continuous Receive / Fallback Operation:} The transceiver must support a low-power continuous receive mode, allowing it to remain active at all times, fitting the role of a \com{} that will be used as an always active fallback, receiving omnidirectionally. This makes a shutdown as a defense infeasible.
  \item \textbf{Obtainable COTS:} We focus on components obtainable COTS by private third parties, excluding vendors that are exclusively selling to government or military customers.
  \item \textbf{Market Relevance:} The \com{} must be relevant in the COTS market. We want to examine representative targets that are of interest to attackers.
  \item \textbf{Business Model Diversity:} The list must include both vendors with closed and open ecosystems, ranging from communication-as-a-service providers to those offering standalone hardware for integration with third-party systems. This diversity enables an evaluation of how different integration models impact product security.
\item \textbf{Security Features:} Each product must incorporate a feature aimed at preventing the issuing of commands from unauthorized parties. We exclude \coms{} that require a different system to handle TC authentication.
  \item \textbf{VHF or UHF  Frequencies:} For each vendor, we focus on VHF/UHF. These bands lower the barrier for adversaries who attempt to build their own ground station and align with the low‑power always‑on requirement.
\end{itemize}

\subsection{Selected Vendors and Components}
For the case study, we selected three \coms{} from three different vendors based on our selection criteria:

    \noindent\textbf{\GomSpace{} \AXonehundred{}}:
A transceiver on the market since 2014 and the only VHF/UHF offering by \GomSpace{}. According to the SatNOGS database, at least 64 active satellites use the \AXonehundred{}, with 174 entries recorded as of July 2025. Our analysis shows this \com{} is deployed across diverse commercial missions, including maritime situational awareness~\cite{eoportalBreizhReconnaissance}, signal intelligence~\cite{eoportalKleosCommercial}, quantum key distribution~\cite{eoportalSpooQy1CubeSat}, and a military-funded ICBM launch detection test mission~\cite{nanosatsERNSTNanosats}. It is also used in orbital transportation services~\cite{skyrocketIONSCV002comma}. Consequently, a vulnerability could, in the worst case, enable the takeover of a transport vehicle and prevent payload deployment, thereby amplifying the impact. Based on public data, \GomSpace{} holds roughly 15\% of the CubeSat COTS component market~\cite{marketsandmarketsCubeSatCompanies} and is consistently ranked among the top vendors~\cite{grandviewresearchCubeSatMarket,lucintelCubeSatMarket,fortunebusinessinsightsCubeSatMarket,imarcgroupCubeSatMarket}.

    \noindent\textbf{\Endurosat{} \UHFtwo{}}:
More than 3000 \Endurosat{} modules are in orbit as of November~2025~\cite{endurosat_homepage}, with over 120 full satellites delivered. The company offers highly integrated solutions, primarily using proprietary protocols while supporting some standards such as \csp{} for the \obc{}. The \UHFtwo{} is its only UHF offering, providing low rate basic communication. This vendor is notable for multi-tenant satellite missions~\cite{skyrocketSharedSat1comma}, where preventing lateral movement is critical, as attackers targeting one payload may compromise others or the \com{}. Like \GomSpace{}, \Endurosat{} is consistently ranked among leading small-satellite vendors~\cite{lucintelCubeSatMarket,grandviewresearchCubeSatMarket,fortunebusinessinsightsCubeSatMarket,imarcgroupCubeSatMarket}.

    \noindent\textbf{\LibreSpace{} \NogsComm{}}:
The \NogsComm{} is a fully open-source hardware \com{} used in past missions and planned for future ones~\cite{curiumsatGitLab,fosdem2025satnogs}. Unlike designs that use separate devices on PC/104 or carrier boards, it integrates both S-Band and UHF transceivers into a single unit. Our analysis is based on the current \NogsComm{} main branch and Curium-1 flight software~\cite{curiumsatGitLab}. We focus on the UHF section, as the S-Band component does not meet our criteria in Section~\ref{sec:com_criteria}. Given the rapid adoption of prior \LibreSpace{} projects such as the SatNOGS \gstation{} network~\cite{satnogs}, especially by cost-sensitive operators, we expect many future designs to build on this module.

\begin{table*}[ht]
\caption{Overview of Vulnerabilities Discovered}
\footnotesize
\begin{threeparttable}
\label{tab:vulnerabilities}

\centering
    \caption{Overview of Vulnerabilities Detected in Each System by Vendor}
\begin{tblr}{
width = \linewidth,
rowhead = 1,
hlines,
colsep  = 0.87pt,
rowsep = 0.0pt,
colspec = {@{} c c c l l c c c @{}},
cell{2}{1} = {r=10}{}, cell{12}{1} = {r=2}{},
cell{14}{1} = {r=2}{},
cell{2}{2} = {r=9}{},  cell{10}{2} = {r=1}{},
cell{12}{2} = {r=2}{},
cell{14}{2} = {r=2}{},
}
Vendor & Segment & Section & Vulnerability (Type\tnote{1})& Attacker (Subcomponent)&  & Exploitable\tnote{2} & Confirmed\tnote{3} \;& Affected Models\tnote{4}\\
\GomSpace{} & Space / Ground & \ref{gom:open_jtag} & Open Debug Port (Impl.)& {\tiny\attackerPA{}} (Debug / Config Interface) &  & \cmark{} & \cmark{} & 2 (AVR32)\\
            & Space / Ground & \ref{gom:replay} & Replay (Design)& {\tiny\attackerAitM{}} (Authentication) &  & \cmark{} & - & 3 (CSP HMAC)\\
            & Space / Ground & \ref{gom:danger_tc} & Dangerous TC (Design)& {\tiny\attackerMI{}} (Internal Commands) &  & \cmark{} & \cmark{} & 3 (CSP)\\
            & Space / Ground & \ref{gom:command_injection} & Command Injection (Impl.)& {\tiny\attackerMI{}} (Network Interfaces) &  & \cmark{} & - & 2 (GOSH \& CSP)\\
            & Space / Ground & \ref{gom:gosh_overflow} & Memory Corruption (Impl.)& {\tiny\attackerMI{}} (Debug / Config Interface) &  & \cmark{} & \cmark{} & 2 (GOSH)\\
            & Space / Ground & \ref{gom:side_channel} & Side Channels (Impl.)& {\tiny\attackerAitM{}} (RF-Frontend, Auth.) &  & ? & ? & -\\
            & Space / Ground & \ref{gom:signature_length} & Insufficient Signature Length (Design)& {\tiny\attackerMT{}} (Authentication) &  & \cmark{} & - & 3 (CSP)\\
            & Space / Ground & \ref{gom:debug_backdoor} & Debug Backdoor (Design + Impl.)& {\tiny\attackerMI{}} (Internal Commands) & & \cmark{} & \cmark{} & 3 (CSP)\\
            & Ground & \ref{gom:swdos} & Syncword Based DoS (Impl.)& {\tiny\attackerMT{}} (Frame Processing) &  & ? & ? & -\\
            & Ground & \ref{sec:satbleed} & \satbleed{} (Impl.)& {\tiny\attackerGSA{}} (Routing) &  & \cmark{} & \cmark{} & 1 (AX100 in GS)\\
\;\;\shortstack{Libre Space \\ Foundation} & Space & \ref{libre:sig_length} & Insufficient Signature Length (Design)& {\tiny\attackerMT{}} (Authentication) &  & \cmark{} & - & 1 \\
            & Space & \ref{libre:spoofing} & TC Authentication Bypass (Design)& {\tiny\attackerAitM{}} (Authentication) &  & \cmark{} & - & 1\\
\Endurosat{} & \;\;\shortstack{Space / Ground \\ (deprecated)}\;\; & \ref{enduro:open_jtag} & Open Debug Port (Impl.)& {\tiny\attackerPA{}} (Debug / Config Interface) &  & \cmark{} & \cmark{} & 2 (MSP430)\\
            & \;\;\shortstack{Space / Ground \\ (deprecated)}\;\; & \ref{enduro:update_integrity} & Insecure Update Service (Design)& {\tiny\attackerMI{}} (Internal Commands) &  & \cmark{} & \cmark{} & 2 (FWUP)
\end{tblr}

\begin{tablenotes}[para]
\scriptsize
\item[1] Design = design flaw; Impl. = implementation flaw.
\item[2] Realistically exploitable?
\item[3] Vendor-confirmed or PoC-demonstrated.
\item[4] Affected models per vendor.
\end{tablenotes}
\end{threeparttable}
\end{table*} To demonstrate the plausibility of the outcomes in Section~\ref{sec:threatmodel} and the applicability of a sub-component-based taxonomy, we conducted a security evaluation with the goal of proving that the outcomes mentioned can be achieved by attackers according to our threat model.
For this, we evaluated the security of the \com{} systems previously identified as critical in Section~\ref{sec:which_coms_are_critical}.
We approached the systems from the perspectives of the attackers introduced in the threat model, investigating flaws in different key aspects.
Where possible, we also looked at the corresponding \fep{} product that is responsible for the terrestrial side of the TT\&C link.
For each \com{} vendor, we first provide context on how this is used and how the protocols are supposed to work to get an intuition about how the vendor expects clients to use the systems. Then we discuss any weaknesses identified during our security evaluation. Finally, we map the vulnerabilities discovered to our threat taxonomy.
You can find an overview of vulnerabilities, including the corresponding attackers and affected models, in Table~\ref{tab:vulnerabilities}.
Appendix Section \ref{appendix:taxonomy} contains data and attack flows for the investigated \coms{}. We also provide vulnerability mappings\footnote{\url{https://github.com/EmsecCispa/SatBleed}}.

\subsection{\GomSpace{} \AXonehundred{}}
\label{subsec:gom}
The \GomSpace{} transceiver is based on FreeRTOS 8.0.1~\cite{amazon2023freertos}, released in 2014~\cite{freertosReleaseHistory}, running on an AVR32 microcontroller and lacks any security hardening such as stack-smashing protection.

The \AXonehundred{} operates as a \csp{} node. \csp{} is a lightweight protocol stack for small satellites, mirroring the Open Systems Interconnection (OSI) model and treating nearly all spacecraft components as network participants~\cite{LibcspLibcsp2024}. It spans multiple physical and data link layers, including I\textsuperscript{2}C, CAN, SPI, UART (RS-232), and Ethernet, enabling deployment across heterogeneous hardware. \GomSpace{} builds all devices around this stack.

CSP routing is static, with each node maintaining a configurable routing table. Subnets often separate spacecraft nodes from ground infrastructure, allowing the \mcs{} to send commands addressed to spacecraft nodes into the network, which are then forwarded to the \cdhs{}. Packets may also be sent directly to subsystems, with intermediaries acting as routers. On multi-device buses such as I\textsuperscript{2}C, \csp{} permits address reuse in underlying protocols~\cite{LibcspLibcsp2024}. Depending on the architecture, the \com{} may act as a node with its own address and services or as an interface for another node (e.g., the \cdhs{}).

The \GomSpace{} \AXonehundred{} uses a custom RF framing mode, \emph{ASM+Golay} (named Mode~5), replacing older, now unsupported modes such as AX.25, to transport CSP packets. It typically functions as a \csp{} node with multiple interfaces and can route traffic (e.g., I\textsuperscript{2}C to CAN). The radio is treated as an interface and can enforce a truncated SHA-1-based HMAC. Packets with missing or incorrect HMACs are discarded regardless of the packet’s HMAC flag.

Unlike libCSP, the \AXonehundred{} does not support XTEA-based encryption or encrypted tunnels to explicitly protect the RF link from unauthorized access. Therefore, RF link encryption cannot be implemented using only the \com{}. An additional device between the PC/104 bus and the \com{} is required, as in the OPS-SAT mission~\cite{willbold2023space}, where the \cdhs{} acts as a gateway.

The \AXonehundred{} provides seven services on ports 0–6, including default libCSP services~\cite{LibcspLibcsp2024}, and exposes them via the RF interface. Thus, even if routed traffic is encrypted, the transceiver’s own services remain externally accessible unless HMAC is enabled. In OPS-SAT~\cite{willbold2023space}, these services were accessible to unauthenticated attackers, meaning that even if the \cdhs{} vulnerability was fixed, the \com{} would remain exposed.

The \GSonehundred{} UHF ground station \fep{} by \GomSpace{} shares the same hardware as the \AXonehundred{} in a rack mount device, and firmware details suggest that it is shared among the ground and space segments. In the following, we expect the \com{} to be used with its official ground station product. We now summarize our findings for the \GomSpace{} \com{}.

\subsubsection{\textbf{Open Debug Port} \attackerPA{}}
\label{gom:open_jtag}
The device has an open JTAG debug interface and serial debug shell exposed on the outside of the enclosure, allowing attackers with short physical access to tamper with the onboard software of the device.

\subsubsection{\textbf{Replay} \attackerAitM{}}
\label{gom:replay}
CSP does not sufficiently protect against the replay of packets. This would have to be enforced by a higher layer in the network stack. Critically, the services on the \com{} itself do not feature additional replay protection. This means that even if the HMAC verification is enabled with a sufficiently long key, an attacker could still replay critical messages, such as changes to the parameters that contain settings.
Note that packets that control the \com{} internal services cannot employ encryption, meaning that an eavesdropper knows exactly what changes a certain packet causes. Then an attacker could compose multiple legitimate changes to parameters and render the mission unable to be commanded effectively, causing persistent \ac{dos}.

\subsubsection{\textbf{Dangerous TC} \attackerMI{}}
\label{gom:danger_tc}
An attacker obtaining a single authenticated \tc{} with their packet payload would already be sufficient to compromise the mission. The attacker could forge a command that reconfigures the \com{} to disable authentication enforcement, effectively bypassing future security checks.
Furthermore, there are many \acp{tc} that can be set to values that lead to persistent \ac{dos}, or are simply usable in a manner that renders the system insecure. Also, the rekeying procedure relies on the \cdhs{} encapsulating the new key in an encrypted message, as commands directly to the \com{} cannot be encrypted. Thus, the only way to securely rekey the \com{} is by sending the new key to the \cdhs{} in an encrypted message, followed by the \cdhs{} setting the key in the \com{} using the internal network interface of the \com{}.

\subsubsection{\textbf{Command Injection} \attackerMI{}}
\label{gom:command_injection}
By redefining the configuration of the \com{}, an attacker can modify onto which UART interface the \emph{GOSH} debug shell is exposed. This enables command injection into said shell, which is intended solely for configuration and debugging of the \com{}. When another \csp{} node is connected to the \com{} via RS-232, any frame sent from that node is interpreted by the \com{} as a debug command. An attacker could exploit this behavior by having a connected node reflect a payload into the debug shell, for example, by using the default \emph{CSP\_PING} service, which echoes back the payload it receives.

\subsubsection{\textbf{Memory Corruption} \attackerMI{}}
\label{gom:gosh_overflow}
The \emph{GOSH} debug shell has the capabilities of sending arbitrary packets and modifying arbitrary parameters and memory. There are also \emph{peek} and \emph{poke} commands that supposedly allow reading and writing memory. The \emph{poke} command in the shell, however, seems not to work in the way we expect. Despite this, an attacker could leverage various debug shell features to gain persistent access or achieve code execution. One example is modifying security-related parameters. The \emph{peek} command contains a memory corruption vulnerability that may also be used to gain code execution.

\subsubsection{\textbf{Timing and RF Leakage Based Side-Channels} \attackerAitM{}}
\label{gom:side_channel}
The \GomSpace{} version of the CSP stack features the same memcmp-based key comparison as in the open-source version libCSP~\cite{libcspComparisonLeaks}. The firmware uses non-constant-time functions for cryptographically sensitive operations such as comparing the result of the HMAC computation to the HMAC field in an incoming packet.
Furthermore, the transmitter appears to be leaking computation-dependent data over RF during transmission. Figure \ref{fig:rf_spectrum_workloads} in the appendix shows that when sending a carrier, the RF emissions differ based on the workload that the processor is currently running, indicating leakage similar to the one described by Camurati~\etal{}~\cite{camurati2018screaming}.
It is unclear whether this is exploitable to further reduce the search space for attacks either attacking the signature of a single packet or the key directly. During reverse engineering, we discovered that every time the firmware needs to compute an HMAC, it retrieves the stored master key and re-runs the key-derivation function rather than deriving the HMAC key once and keeping it in memory for reuse. This design both wastes processing cycles and expands the window for potential side-channel leaks.

\subsubsection{\textbf{Insufficient Signature Length} \attackerMT{}}
\label{gom:signature_length}
The SHA1-HMAC employed in \csp{} is truncated to a 32-bit length. On slow links, performing an exhaustive online brute-force attack is generally infeasible due to the low data rate. However, the protocol is also supported on high data-rate transceivers that operate in the multi megabit per second data-rates~\cite{gomspaceLinkSX}, which makes sending 2\textsuperscript{32} packets a feasible attack. This becomes particularly feasible if an entire constellation shares the same key, allowing guesses to be balanced across many satellites and their ground stations. Assuming a malicious ground station is able to achieve a data rate of 50 Mbit/s, an online attack to forge the authentication tag could take under 10 hours.

\subsubsection{\textbf{Debug Protocol Allowing Arbitrary Memory Write} \attackerMI{}}
\label{gom:debug_backdoor}
The \emph{CMP} service defined in the open-source libCSP allows arbitrary memory write. However, this functionality shall be disabled or replaced by weaker functionality. In the implementation, it is explicitly marked as extremely dangerous~\cite{LibcspLibcsp2024}. The \GomSpace{} implementation of the \emph{CMP} service also has this feature; it uses a function that allows access to parameter memory. However, the memory address of the function pointer that gets called to do the memory access is also accessible via this service. This means that an attacker could use this to gain arbitrary code execution or create an arbitrary memory read/write primitive. The vendor did not confirm that this is on purpose and instead answered that this protocol is not supposed to be used. This means that an attacker who can have data processed by the \com{} can gain code execution.

\subsubsection{\textbf{Syncword-Based DoS} \attackerMT{}}
\label{gom:swdos}
In response to the unusually complex sync-word detection approach, likely intended to extend the capabilities of the off-the-shelf RF IC, we developed a fuzzer, \fuzzername{}, targeting physical layer protocol bugs. The tool generates streams of symbols containing sync-words and occasional valid frames for the system to process. Rather than merely using the RF layer to feed higher protocol layers, it produces inputs designed to trigger faults in the lowest level of frame handling.

While running \fuzzername{}, we observed error messages related to RF chip and MCU communication. Over time, the setup triggered a condition where the system continuously received frames at what appeared to be its maximum rate, even without a transmitter. This behavior is distinct from rare idle cases where random noise produces spurious frames.

The effect is likely caused by a race condition that misassigns responsibility for sync-word detection. Under normal operation, the radio IC detects the sync-word and interrupts the MCU, which then handles detection for subsequent frames in software. However, during fuzzing, the \com{} intermittently stops waiting for a sync-word and treats all incoming data as valid frames. This likely occurs because both the software and RF chip enter a state where each assumes the other is responsible for detection, causing all RF input to be accepted as frames. Since these spuriously detected frames contain random payloads without valid length fields or signatures, they would go unnoticed in practice, aside from degraded reception due to incorrect framing.

This “syncword bug” appears independent of specific physical layer inputs, as timing-accurate replays do not reliably trigger it. Instead, it likely depends on internal timing conditions combined with certain input sequences. This behavior highlights how unconventional transceiver designs can introduce a new class of unauthenticated DoS vulnerabilities.

\subsubsection{\textbf{Logic Error in TM Handling (\satbleed{})} \attackerAitM{}}
\label{sec:satbleed}
\begin{figure}[t]
\centering
\footnotesize
\resizebox{1.0\columnwidth}{!}{\begin{tikzpicture}[
  actor/.style={draw,rounded corners=3pt, rectangle, minimum width=2cm, minimum height=.8cm, align=center},
  lifeline/.style={gray, dashed},
  msg/.style={-Stealth, thick},
  keymsg/.style={-Stealth, thick},
  >=Stealth
]
\begin{scope}[yscale=0.79]

\node[actor]                    (sat) {Satellite};
\node[actor, right=2cm of sat]  (att) {Attacker};
\node[actor, right=2cm of att]  (gs)  {Ground\\Station};

\foreach \n in {sat,att,gs}
  \draw[lifeline] (\n.south) -- ++(0,-9.4);

\newcommand{\at}[2]{($(#1.south)+(0,#2)$)}

\begin{scope}[on background layer]
\coordinate (boxSW) at ($(sat.south west)+(0,-1.9)$);
  \coordinate (boxNE) at ($(gs.south east)+(0,-4.2)$);
  \path[draw=black!60, thick, fill=black!10, rounded corners=2pt]
       (boxSW) rectangle (boxNE);
  \node[font=\footnotesize\bfseries, anchor=south west]
       at ($(boxSW)+(0.00,-0.60)$) {Attacker Injects Malicious TM };
\end{scope}

\draw[msg] \at{sat}{-0.8} -- \at{gs}{-0.8}
           node[midway, above] {TM\textsubscript{Beacon}};
\draw[msg] \at{sat}{-1.6} -- \at{gs}{-1.6}
           node[midway, above] {TM\textsubscript{Beacon}};

\draw[msg]    \at{att}{-2.4} -- \at{gs}{-2.4}
           node[midway, above] {Malicious TM};

\draw[keymsg] \at{gs}{-3.2} -- \at{sat}{-3.2}
           node[midway, above] {$K \;\|\; \mathrm{HMAC}_{k}(K)$};
\draw[keymsg] \at{gs}{-4.0} -- \at{att}{-4.0}
           node[midway, above] {$K \;\|\; \mathrm{HMAC}_{k}(K)$};

\draw[msg] \at{sat}{-4.8} -- \at{gs}{-4.8}
           node[midway, above] {TM\textsubscript{Beacon}};

\draw[dashed, thin, gray]
      ($(sat.south)+(0,-5.3)$) -- ($(gs.south)+(0,-5.3)$)
      node[midway, below=2pt] {\scriptsize TM Signature enforced after Acquisition of Signal};

\draw[keymsg] \at{att}{-6.3} -- \at{sat}{-6.3}
           node[midway, above] {$\text{TC}_{Mal} \;\|\; \mathrm{HMAC}_{k}(\text{TC}_{Mal})$};
\draw[keymsg] \at{att}{-7.1} -- \at{gs}{-7.1}
           node[midway, above] {$\text{TC}_{Mal} \;\|\; \mathrm{HMAC}_{k}(\text{TC}_{Mal})$};

\draw[keymsg] \at{gs}{-7.9} -- \at{sat}{-7.9}
           node[midway, above] {$\text{TC} \;\|\; \mathrm{HMAC}_{k}(\text{TC})$};
\draw[keymsg] \at{sat}{-8.7} -- \at{gs}{-8.7}
           node[midway, above] {$\text{TM} \;\|\; \mathrm{HMAC}_{k}(\text{TM})$};

\end{scope}
\end{tikzpicture}
}\caption{
Sequence diagram of an attacker performing a \satbleed{} type attack.
\tm{} injection into the ground station while it waits for the beacons of the satellite. The line after which TM signatures are enforced may not exist, depending on the mission; however, if a switch is present, it may occur after beacons have been received. }
\label{fig:beacon_attack}
\end{figure}
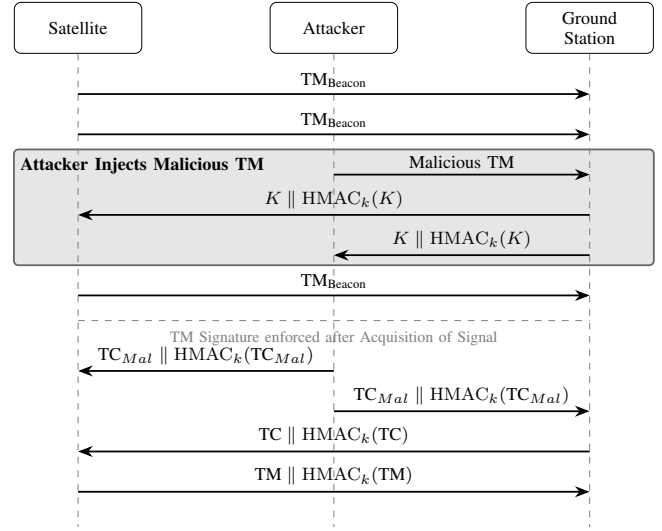
 \GomSpace{} utilizes the same module in both the satellite and the ground station device, with identical functionality in the ground and space versions of the \AXonehundred{}. However, these two deployment scenarios should not necessarily behave identically. In particular, the ground version exposes the same TC interface over RF as the space version. As a result, the most critical attack we found, \satbleed{} (described in Figure \ref{fig:beacon_attack}), becomes possible. An attacker can inject a parameter request into the ground station even though the ground station has not sent a corresponding command. In our attack, the injected request asks for the key used to sign TC. In principle, the ground station should discard this nonsensical request of the spacecraft, which is effectively the spacecraft asking the ground station for its own key, yet a response is nevertheless sent. This should not be understood as merely a consequence of missions disabling signatures. In practice, signatures or encryption may be enforced at a higher layer, with the cryptographic processing equipment placed behind the FEP. In such a setup, the FEP would normally not interpret \tm{}, since, unlike the \com{}, there is no need to expose a command interface for telemetry over the RF interface, as such an interface is only needed for recovery purposes of the \com{}. Consequently, the vulnerability persists even in the presence of encrypted \tm{}. We further discuss the impact of \satbleed{} in Section \ref{sec:satnogs_hmac_analysis}.

Finally, we noticed that the \AXonehundred{} uses Carrier Sense Multiple Access, and does not start transmitting if it detects an RSSI greater than -95dbm by default, making jamming an effective measure as it will cease transmitting when a low-power signal is received by it.

\subsection{\LibreSpace{} \NogsComm{}}
The \NogsComm{} is a \com{} by \LibreSpace{}. The product is a unified UHF and S-Band hardware. It utilizes a combination of an Xilinx Zynq Cortex A9 SoC and an Arm Cortex M7 based microcontroller that is always powered on.

\subsubsection{\textbf{Insufficient Signature Length} \attackerMT{}}
\label{libre:sig_length}
The \NogsComm{} uses a \tc{} protocol based on Protobuf~\cite{satnogs_protobuf}. When transmitting the same commands over RF, a 4-byte authentication tag is added to each packet. Similar to Vulnerability \ref{gom:signature_length}, the length of this tag may not be sufficient.
On top of this framing layer, a mission can add whichever protocols it desires. One of the missions continues by defining commands as single-frame commands for basic operations. It seems plausible that a more complex mission adapts protocols such as \csp{}.

\subsubsection{\textbf{Authentication Bypass by Spoofing} \attackerAitM{}}
\label{libre:spoofing}
The authentication for the \NogsComm{} is based on a secret that is appended to a TC when transmitted over RF. The secret is then hashed and compared against a stored result. In case the check passes, the frame is processed. This means that if the attacker observes this secret once, they can append it to any payload and have it be authenticated.
The key does not change, so a passive eavesdropper can intercept a single TC and use that to send arbitrary \acp{tc}. However, this does not mean an attacker can gain persistence, as Over-the-Air updates are cryptographically verified.

\subsection{\Endurosat{} \UHFtwo{}}
From \Endurosat{} we analyzed the \UHFtwo~transceiver.
Unlike the open protocols \csp{} and the protocol used by the \NogsComm{}, \Endurosat{} uses the so-called \esps{}. It is a proprietary protocol stack that lacks an open-source implementation and has no publicly accessible documentation. There do not seem to be any other vendors besides \Endurosat{} using it. Some layers of \esps{} can be used to transport \csp{} to ensure compatibility.

The \UHFtwo{} has two modes of operation: in one, it behaves like a transparent RF interface, while in the other, it functions as a network node that can be configured over the air.
When \esps{} data is transported over the radio, there is a special encapsulation layer that carries encrypted data and thus does link encryption.
Similar to the \csp{}, there is an address space of addresses that are used for devices on the network. Also, there is an initial broadcast when joining a network.

After manually reverse engineering the device's circuits and firmware, we chose partial re-hosting for fuzzing the firmware. Additionally, we reverse-engineered an undocumented update protocol, as this system, or a similar one, is used in multi-payload missions where an attacker on an internal data bus is a plausible threat.

\subsubsection{\textbf{Open Debug Port} \attackerPA{}}
\label{enduro:open_jtag}
The Spy-Bi-Wire debug and serial bootloader of the microcontroller is accessible without authentication. This allows attackers with short-term access to manipulate firmware. Afterward, there is no capability of restoring the firmware to a known good state.

\subsubsection{\textbf{Firmware Update Without Integrity Check} \attackerMI{}}
\label{enduro:update_integrity}
The firmware updates are not cryptographically signed and thus do not protect against a compromised device on the bus pushing a malicious update. Note that updating the software does not imply resetting the device's configuration. Encryption keys stored on the device may be exposed to attackers after gaining code execution through this protocol, sidestepping the limitation of key material being a write-only configuration.
Software components also suggest the existence of a second undocumented protocol, which we did not investigate further, as we have already demonstrated that a local attacker can gain code execution.
\section{Feasibility and Impact of \satbleed{}}
\label{sec:satnogs_hmac_analysis}

To better understand which active systems may be affected by the critical \satbleed{} vulnerability described in Section~\ref{sec:satbleed}, we estimate how many missions using this \com{} transmit telemetry beacons in configurations where the GS100 cannot enforce an HMAC field. It is important to distinguish between the existence of the \satbleed{} vulnerability, which we have conclusively confirmed, and its impact on active missions, which can only be inferred heuristically.

Because no public information exists on how operators configure and deploy the \AXonehundred{}, we rely on publicly available observation data from the SatNOGS project~\cite{satnogs}. This open-source dataset includes telemetry beacon recordings from 80 transmitters using the GomSpace proprietary Mode~5 framing with FSK/MSK modulation, corresponding to 64 active missions as of August~2025. We infer the use of the \AXonehundred{} because it is the only known device implementing this mode, and thus attribute these transmissions to missions using this \com{}, which we decode and analyze as detailed in Appendix~\ref{appendix:satnogs_details}.

Our analysis is based on several assumptions. First, we assume in-orbit \acp{com} and \acp{fep} run firmware versions equivalent to those available in April~2025, in which we identified the vulnerability. Second, we assume the vendor does not provide mission-specific hardware or firmware variants, as we found no evidence of such deviations. Third, we assume the ground infrastructure is vendor-provided and plausibly based on the GS100, representing the simplest deployment model.

However, beacon observations do not reveal all configuration parameters. They do not indicate whether encryption or authentication is applied to all telemetry, nor do they prove the presence or absence of signatures on non-beacon traffic. Instead, they only reflect which ground-station configuration is active for a mission at a given time, which is still sufficient to infer security-relevant behavior.

Under these assumptions, 28 of the 64 active missions in the dataset are plausibly vulnerable to \satbleed{}. Of these, 12 transmit beacons without GS100-enforceable HMAC protection, while 16 frequently toggle this configuration.

This methodology has inherent limitations. Most importantly, it relies entirely on passive observation rather than active validation, so the result should be interpreted as a lower bound, with the true number likely higher. Some missions use encrypted or compressed beacons, complicating reliable classification. Moreover, telemetry encryption does not prevent \satbleed{}, as the \com{} processes commands before forwarding them to subsystems that may perform decryption. Finally, our analysis is limited to missions observed by the SatNOGS network.

\section{Contextual Factors Behind Security Gaps}

\subsubsection*{Testing Challenges}
Security testing of \coms{} is inherently difficult. The radios tightly integrate hardware and software and must remain operational in orbit despite progressive degradation. Firmware therefore includes extensive coherence and self-diagnostic mechanisms, as partial subsystem failures are expected due to cumulative ionizing dose over time. These constraints limit automated partial re-hosting approaches, where the CPU is emulated without peripherals.

When firmware continuously polls peripherals and timing-critical tasks are shared between the RF front-end and MCU, such as sync-word detection handover, accurate emulation requires precise hardware models unavailable outside the physical device. Even with such models, many radio subsystems use exotic processor architectures lacking fully functional public instruction set emulators.

Space communication subsystems also employ protocols that diverge from commercial standards. As a result, firmware operates on lower-level data than typical Internet of Things (IoT) systems, processing symbol streams and reconstructing frame boundaries or sequencing. Thus, re-hosting fuzzers struggle to reach deeper protocol layers without significant manual effort. Consequently, most security testing must be conducted on flight-representative hardware, which is costly and time-consuming.

\subsubsection*{No Separation of Data and Commands Between Systems}
Across all evaluated systems, there is no explicit separation between user data transported by the \com{} and management commands that bypass the nominal \cdhs{} path and directly address the \com{} subsystem. The absence of authentication per purpose implies that any entity capable of injecting traffic on the radio link, including payloads unrelated to platform control, can issue commands to the transceiver and potentially seize control. This is reflected in the data flow diagrams, as none of the \coms{} required data flows to the Internal Command Interface from an internal interface to pass authentication.

\subsubsection*{Missing Secure Boot Features}
Of the three \coms{} evaluated, two provide no mechanism that assures software integrity at boot time. Only the \NogsComm{} enforces secure boot through cryptographic signatures.

\subsubsection*{Non-Defensive Use of Cryptographic Libraries}
During analysis, we recognized that the \NogsComm{} and \AXonehundred{} are not following the defensive usage of crypto-related code. They are using non-constant-time comparisons. The \NogsComm{} is discarding error values of crypto functions, just assuming that operations were successful. In the space environment, this is especially critical, as high-energy particle-induced faults may leave programs in an invalid state, making defensive error checking especially important.

\section{Security Hardening Measures}
Moving satellite components in the direction of zero trust is highly important, moving on from architecting satellites to act like a homogeneous system from a security standpoint to systems no longer trusting each other. As a first step in this direction, signatures may be checked on the device acting on the command instead of a device that is forwarding commands. For example, an \ocs{} could feature authenticated thruster commands that are only valid once or when executed in a certain time period. This could prevent malware on other components from masquerading as a \cdhs{} and having the satellite perform an orbital maneuver.

\subsubsection*{Introduction of Privilege Separation}
One of the issues that all the systems have is that they lack a clear distinction between data that the systems transport and commands that bypass the normal \cdhs{} and can be issued directly to the \com{} instead. Introduction of authenticated internal commands, as shown in Figure \ref{fig:taxonomy}, is vital.
\subsubsection*{Moving Away From PC/104}
By the very working principle of the PC/104 connector, it makes sense to have many subsystems and components sharing the same digital interfaces for communication. The issue is that the components for space systems are usually supported for extended time periods and thus only adopt attack prevention measures slowly. Especially, as for example, new CAN transceivers that prevent attacks will have to be evaluated for successful operation in conditions that they have not been designed for.
We suggest a novel bus interface based on optical communication, which maintains the advantages of a shared medium while enabling the isolation of individual nodes by wavelength. Additionally, this approach provides galvanic isolation~\cite{palmer2025wireless}.

\subsubsection*{Recovery Methods for On-board Software}
In typical systems, compromises can often be resolved by redeploying software to restore a known secure state. This approach is not viable for satellites, as even fallback software cannot be assumed to be trustworthy. Therefore, a supervisory system is needed that can be attested as not compromised and capable of restoring the satellite’s state. Such a system introduces challenges, as it must not increase failure risk and must, by design, be able to override core functionality.

\subsubsection*{Adapting Best Practices from Other Domains}
A major issue in \coms{} is the software monoculture in space. In the early Internet, many publicly accessible systems ran identical software and were frequently exploited. A similar problem is likely for satellites. Experience from the Internet shows that rapid patching and defense-in-depth mechanisms are effective. However, patches are rarely deployed in space hardware due to safety concerns, so mechanisms are needed to apply them safely. Techniques that guarantee patch properties, such as~\cite{2022HeRapidPatch}, could shift \cots{} vendor practices and reduce the exposure time of known vulnerabilities. Furthermore, as shown in Section~\ref{sec:case_study}, even basic mitigations like stack canaries are not widely used in space components. Vendors should reassess whether the risk of operational issues from such mitigations outweighs the security benefits. The authors of \cite{jero2024securing} also outline a path toward a more secure and resilient software stack.

\subsubsection*{Restrictions Posed by the Space Environment}

Because mission failure is extremely costly, operators may hesitate to deploy security mechanisms that increase power use, add complexity, or introduce new failure points. This is especially true for mitigations requiring additional system states or computation. In such cases, watchdog timers can disable security features in a controlled way after access is lost due to an attack or failure. This allows operators to decide whether to passivate after losing a security function or continue operations without it and accept the risk.

More generally, security mechanisms should support phased deployment, enabling rollout to part of a system before extending protection to all components. As in conventional supervisory systems, security features also need backups or multiple ways to achieve the same goal, though such redundancy must be carefully hardened.

Finally, proposed mitigations must be evaluated against space environment constraints, including resilience to radiation-induced single-event upsets (SEUs) and potential impacts on component lifetime due to increased processing or resource use.

\section{Discussion}

\subsubsection*{Other Affected \coms{}}
Upon disclosure, the affected vendors confirmed to us that other \coms{} in their portfolio are also affected by some of the vulnerabilities. This means that the total number of impacted missions probably significantly overshadows assumptions based on the use of specific hardware.

\subsection{Limitations}

\subsubsection*{Open Source Telemetry Survey}
When analyzing the impact of \satbleed{} in Section \ref{sec:satnogs_hmac_analysis}, we identify which satellite missions use HMACs in their beacons. Due to the lack of disclosure from mission operators, we rely on indirect indicators, namely the entropy of suspected HMAC fields and the frequency of HMAC flags in the payloads.
To avoid false positives caused by bit errors in the flags field, we also confirm the presence of the HMAC field in the header by examining the entropy of that field. Note again that this approach does not work if the mission employs encryption/compression, as the payload would also have high entropy, making it impossible to detect the HMAC field directly before the payload. Therefore, we deliberately exclude missions employing encrypted/compressed telemetry. Importantly, neither mitigates \satbleed{} but both obscure the presence of integrity protection. As a result, this exclusion systematically omits missions that may still be vulnerable, ensuring that our findings constitute a strict lower bound on the true impact of \satbleed{}.

\subsubsection*{Representativeness of \com{} Samples}
Although we analyzed major providers with a significant market share, the NewSpace ecosystem is highly diverse. Mission architectures, protocol stacks, and deployment models vary widely across systems. While the subcomponent-based perspective and the resulting threat taxonomy are designed to be generalizable, the concrete issues and vulnerabilities we uncovered are not necessarily so, since some arise from provider-specific implementation choices. Our findings therefore indicate that security oversights are common, but they do not imply that every system is likely vulnerable.

\subsubsection*{Completeness of Methodology}
Our vulnerability discovery process relied on manual reverse engineering, physical hardware inspection, and targeted fuzzing. We do not prove the absence of further flaws within these devices. Given the lack of transparency in the firmware, it is likely that additional vulnerabilities exist within the analyzed \coms{}.
\section{Conclusion}
In this paper, we develop the first comprehensive threat model for \coms{} and apply it to an in-depth analysis of communications subsystems from three major vendors. Our evaluation uncovers multiple critical vulnerabilities, some of which, when traced through open-source SatNOGS telemetry, reveal 28 vulnerable in-orbit active satellites. Beyond this, our broader testing uncovered systemic weaknesses, ranging from insufficient replay defenses to insecure update mechanisms that collectively expose the unprotected nature of current \com{} modules.
Overall, our work highlights the urgent need to embed security throughout the design lifecycle of \com{} subsystems. Without such an approach, small satellites will remain vulnerable, undermining their operational reliability and hindering the advancement of resilient space assets.
\section{Ethics considerations}
\subsubsection*{Responsible Disclosure}
We adhere to responsible disclosure principles and have responsibly disclosed all identified vulnerabilities to their vendors. \GomSpace{} and \Endurosat{} have both acknowledged the reported vulnerabilities and plan to address them in the next generation of their communication modules. \GomSpace{} additionally stated that fixes are also planned for some of their currently available models.
Furthermore, we are engaged in discussions with federal institutions to address the vulnerabilities and, in very critical cases, notify the affected missions directly.
\subsubsection*{Non-Attribution of Vulnerable Missions}
Identifying active spacecraft that operate on these vulnerable \com{} systems, enables malicious actors to target them using the information in our publication. To mitigate this risk, we only present summarized results and statistical analysis, without linking vulnerabilities to individual mission names or operators. Although it is relatively trivial to derive the identity of the missions based on public data, we have decided not to publish the list of the 28 vulnerable missions.
\subsubsection*{Withholding of Exploit Payloads and Proof-of-Concept Code}
No working exploits, payload generators are released as part of this publication. While technical details are described to a certain level, supporting academic reproducibility, we intentionally limit them to prevent immediate weaponization of our work.
\subsubsection*{Avoiding Harm During Analysis}
All experimental evaluations were performed on isolated test setups or on vendor-provided hardware in a controlled environment. We not only avoided transmitting on RF bands, but also took strict precautions to ensure there was no RF leakage from our test setups. Additionally, we did not interact with the communication links of any operational satellite.
\subsubsection*{Ethical Limitations}
Despite our best efforts, we are aware that any publication addressing vulnerabilities in widely used critical infrastructure may carry a certain risk. Attackers with sufficient motivation and technical expertise may be able to replicate or infer parts of our work. However, the benefits of disclosure, vendor response, and public awareness outweigh these risks.

\section*{Acknowledgments}
The project on which this report is based was supported with funding from the German Federal Ministry of Research, Technology and Space (BMFTR) under the grant number 50YB2607B. The author is responsible for the content of this publication. We further thank the Saarbrücken Graduate School of Computer Science for their funding and support.

\section{LLM usage considerations}
LLMs were used for editorial purposes in this manuscript, and all outputs were inspected by the authors to ensure accuracy and originality.

\bibliographystyle{IEEEtran}
\bibliography{IEEEabrv,bibl}

\begin{thebibliography}{10}
\providecommand{\url}[1]{#1}
\csname url@samestyle\endcsname
\providecommand{\newblock}{\relax}
\providecommand{\bibinfo}[2]{#2}
\providecommand{\BIBentrySTDinterwordspacing}{\spaceskip=0pt\relax}
\providecommand{\BIBentryALTinterwordstretchfactor}{4}
\providecommand{\BIBentryALTinterwordspacing}{\spaceskip=\fontdimen2\font plus
\BIBentryALTinterwordstretchfactor\fontdimen3\font minus
  \fontdimen4\font\relax}
\providecommand{\BIBforeignlanguage}[2]{{%
\expandafter\ifx\csname l@#1\endcsname\relax
\typeout{** WARNING: IEEEtran.bst: No hyphenation pattern has been}%
\typeout{** loaded for the language `#1'. Using the pattern for}%
\typeout{** the default language instead.}%
\else
\language=\csname l@#1\endcsname
\fi
#2}}
\providecommand{\BIBdecl}{\relax}
\BIBdecl

\bibitem{manulis2021cyber}
M.~Manulis, C.~P. Bridges, R.~Harrison, V.~Sekar, and A.~Davis, ``Cyber
  security in new space: Analysis of threats, key enabling technologies and
  challenges,'' \emph{International Journal of Information Security}, vol.~20,
  no.~3, pp. 287--311, 2021.

\bibitem{attack_on_viasat}
N.~Boschetti, N.~G. Gordon, and G.~Falco, ``Space cybersecurity lessons learned
  from the viasat cyberattack,'' in \emph{ASCEND 2022}, 2022, p. 4380.

\bibitem{saalman_saveljeva2023mapping}
\BIBentryALTinterwordspacing
L.~Saalman, L.~Saveleva~Dovgal, and F.~Su, ``Mapping cyber-related missile and
  satellite incidents and confidence-building measures,'' Stockholm
  International Peace Research Institute (SIPRI), SIPRI Insights on Peace and
  Security No. 2023/10, Nov. 2023, accessed \today. [Online]. Available:
  \url{https://www.sipri.org/sites/default/files/2023-11/2023\_10\_cyber\_mapping\_incidents.pdf}
\BIBentrySTDinterwordspacing

\bibitem{symantec_thrip_2018}
{Symantec Threat Hunter Team}, ``Thrip: Espionage group hits satellite,
  telecoms, and defense companies,''
  \url{https://www.security.com/threat-intelligence/thrip-hits-satellite-telecoms-defense-targets},
  Jun. 2018, threat Intelligence blog post; 5min read.

\bibitem{flaherty2014chinese}
M.~P. Flaherty, J.~Samenow, and L.~Rein, ``Chinese hack u.s. weather systems,
  satellite network,'' \emph{The Washington Post}, accessed on \today.

\bibitem{willbold2023space}
J.~Willbold, M.~Schloegel, M.~Vögele, M.~Gerhardt, T.~Holz, and A.~Abbasi,
  ``Space odyssey: An experimental software security analysis of satellites,''
  in \emph{IEEE Symposium on Security and Privacy (SP)}, 2023, pp. 1--19.

\bibitem{scharnowski2023case}
T.~Scharnowski, F.~Buchmann, S.~W{\"o}rner, and T.~Holz, ``A case study on
  fuzzing satellite firmware,'' in \emph{Workshop on the Security of Space and
  Satellite Systems (SpaceSec)}, 2023.

\bibitem{peled2023evaluating}
R.~Peled, E.~Aizikovich, E.~Habler, Y.~Elovici, and A.~Shabtai, ``Evaluating
  the security of satellite systems,'' \emph{arXiv preprint arXiv:2312.01330},
  2023.

\bibitem{falco2020sat2sat}
G.~Falco, ``When satellites attack: Satellite-to-satellite cyber attack,
  defense and resilience,'' in \emph{Proceedings of the AIAA ASCEND Conference:
  Accelerating Space Commerce, Exploration, and New Discovery}.\hskip 1em plus
  0.5em minus 0.4em\relax American Institute of Aeronautics and Astronautics
  (AIAA), 2020, p. 4014.

\bibitem{pavur2019cyberasat}
J.~Pavur and I.~Martinovic, ``The cyber-asat: On the impact of cyber weapons in
  outer space,'' in \emph{International Conference on Cyber Conflict (CyCon)},
  vol. 900, 2019, pp. 1--18.

\bibitem{rawlins2022death}
F.~Rawlins, R.~Baker, and I.~Martinovic, ``Death by a thousand cots: Disrupting
  satellite communications using low earth orbit constellations,'' in
  \emph{Workshop on the Security of Space and Satellite Systems (SpaceSec)},
  2023.

\bibitem{pavur2020ataleof}
J.~Pavur, D.~Moser, M.~Strohmeier, V.~Lenders, and I.~Martinovic, ``A tale of
  sea and sky on the security of maritime vsat communications,'' in \emph{IEEE
  Symposium on Security and Privacy (SP)}, 2020, pp. 1384--1400.

\bibitem{willbold2024vsaster}
J.~Willbold, M.~Schloegel, R.~Bisping, M.~Strohmeier, T.~Holz, and V.~Lenders,
  ``Vsaster: Uncovering inherent security issues in current vsat system
  practices,'' in \emph{ACM Conference on Security and Privacy in Wireless and
  Mobile Networks (WiSec)}, 2024, pp. 288--299.

\bibitem{satnogs}
{Libre Space Foundation / SatNOGS Project}, ``Satnogs – satellite networked
  open ground station,'' \url{https://www.satnogs.org/}, 2025, open‑source
  global network of satellite ground stations. Initial release April 2014.

\bibitem{nano2025howmanysats}
``How many satellites are in space?''
  \url{https://nanoavionics.com/blog/how-many-satellites-are-in-space/}, 2025,
  [Accessed 26-05-2025].

\bibitem{nanosatsFigures}
\BIBentryALTinterwordspacing
E.~Kulu, ``Nanosats database,'' 2025, [Accessed 30-04-2025]. [Online].
  Available: \url{https://www.nanosats.eu/\#figures}
\BIBentrySTDinterwordspacing

\bibitem{tinygs}
``tinygs-webapp,'' \url{https://tinygs.com/}, [Accessed 21-07-2025].

\bibitem{amazon2023freertos}
\BIBentryALTinterwordspacing
{Amazon Web Services}, ``Freertos - market leading rtos (real time operating
  system) for embedded systems with internet of things extensions,'' Dec. 2023.
  [Online]. Available: \url{https://www.freertos.org/index.html}
\BIBentrySTDinterwordspacing

\bibitem{zephyrRtos}
{Zephyr Project, Linux Foundation}, ``Zephyr project – scalable real‑time
  operating system,'' \url{https://www.zephyrproject.org/}, 2025, open‑source
  RTOS hosted by the Linux Foundation. Release v4.1.0 (March 7, 2025); supports
  multi‑architecture embedded systems.

\bibitem{rtems}
{RTEMS Project, OAR Corporation}, ``Rtems – real‑time executive for
  multiprocessor systems,'' \url{https://www.rtems.org/}, 2025, open‑source
  real‑time operating system for embedded and space applications; initial
  release 1993; latest stable release 6.1 on 2025‑01‑22.

\bibitem{AlenSpaceTrevo}
{Alén Space}, ``{T}{R}{E}{V}{O}: {M}odular {H}igh {P}erformance {S}{D}{R}
  {P}latform | {A}lén {S}pace --- alen.space,''
  \url{https://alen.space/products/trevo/}, highly flexible Software Defined
  Radio (SDR) with flight heritage.

\bibitem{rederlechner2026one}
J.~Rederlechner, U.~Planta, and A.~Abbasi, ``One small patch for a file, one
  giant leap for ota updates,'' 2026.

\bibitem{mccomas2016core}
D.~McComas, J.~Wilmot, and A.~Cudmore, ``The core flight system (cfs)
  community: Providing low cost solutions for small spacecraft,'' in
  \emph{AIAA/USU Conference on Small Satellites}.\hskip 1em plus 0.5em minus
  0.4em\relax Logan, UT: {Utah State University}, {University Libraries}, 2016.

\bibitem{Bocchino2018FPrime}
\BIBentryALTinterwordspacing
R.~Bocchino, T.~Canham, G.~Watney, L.~Reder, and J.~Levison, ``F prime: An
  open-source framework for small-scale flight software systems,'' in
  \emph{AIAA/USU Conference on Small Satellites}, Logan, UT, 2018. [Online].
  Available: \url{https://digitalcommons.usu.edu/smallsat/2018/all2018/328/}
\BIBentrySTDinterwordspacing

\bibitem{raccoonos}
P.~W. José Manual~Diez, Fabian~Krech, ``Raccoon os,''
  \url{https://gitlab.com/rccn}, [Accessed 20-01-2025].

\bibitem{magnetorquer}
\BIBentryALTinterwordspacing
L.~J. Kamm, ``Magnetorquer-a satellite orientation device,'' \emph{ARS
  Journal}, vol.~31, pp. 813--815, 1961. [Online]. Available:
  \url{https://api.semanticscholar.org/CorpusID:122456319}
\BIBentrySTDinterwordspacing

\bibitem{esa_tmtc}
{European Space Agency}, ``{Telemetry \& Telecommand – Onboard Computers \&
  Data Handling},''
  \url{https://www.esa.int/Enabling\_Support/Space\_Engineering\_Technology/Onboard\_Computers\_and\_Data\_Handling/Telemetry\_Telecommand},
  accessed: 2025-07-15.

\bibitem{tedeschi2022satellite}
P.~Tedeschi, S.~Sciancalepore, and R.~Di~Pietro, ``Satellite-based
  communications security: A survey of threats, solutions, and research
  challenges,'' \emph{Computer Networks}, vol. 216, p. 109246, 2022.

\bibitem{driouch2024distids}
O.~Driouch, S.~Bah, and Z.~Guennoun, ``Distributed intrusion detection system
  for cubesats, based on deep learning packets classification model,'' in
  \emph{2024 Security for Space Systems (3S)}, 2024, pp. 1--8.

\bibitem{he2019security}
D.~He, X.~Li, S.~Chan, J.~Gao, and M.~Guizani, ``Security analysis of a
  space-based wireless network,'' \emph{IEEE Network}, vol.~33, no.~1, pp.
  36--43, 2019.

\bibitem{jero2024securing}
S.~Jero, J.~Furgala, M.~A. Heller, B.~Nahill, S.~Mergendahl, and R.~Skowyra,
  ``Securing the satellite software stack,'' in \emph{Workshop on the Security
  of Space and Satellite Systems (SpaceSec)}, 2024.

\bibitem{deadf00dHackedESAs}
D.~Maurice-Michel, ``{H}ow {I} hacked an {E}{S}{A}'s experimental satellite ---
  deadf00d.com,''
  \url{https://www.deadf00d.com/post/how-to-hack-an-esa-experimental-satellite.html},
  [Accessed 09-07-2025].

\bibitem{smailes2023waththis}
\BIBentryALTinterwordspacing
J.~Smailes, S.~K\"{o}hler, S.~Birnbach, M.~Strohmeier, and I.~Martinovic,
  ``Watch this space: Securing satellite communication through resilient
  transmitter fingerprinting,'' in \emph{ACM Conference on Computer and
  Communications Security (CCS)}, ser. CCS '23.\hskip 1em plus 0.5em minus
  0.4em\relax New York, NY, USA: Association for Computing Machinery, 2023, p.
  608–621. [Online]. Available: \url{https://doi.org/10.1145/3576915.3623135}
\BIBentrySTDinterwordspacing

\bibitem{wigchert2025detection}
J.~Wigchert, S.~Sciancalepore, and G.~Oligeri, ``Detection of aerial spoofing
  attacks to leo satellite systems via deep learning,'' \emph{Computer
  Networks}, p. 111408, 2025.

\bibitem{masson2024sdlsimpl}
L.~Masson, M.~Bonjour, L.~Thoeny, and S.~Willy, ``Developing a ccsds compliant
  platform to reliably secure current and future space data links,'' in
  \emph{2024 Security for Space Systems (3S)}, 2024, pp. 1--8.

\bibitem{willbold2024satellite}
J.~Willbold, F.~Sciberras, M.~Strohmeier, and V.~Lenders, ``Satellite
  cybersecurity reconnaissance: Strategies and their real-world evaluation,''
  in \emph{IEEE Aerospace Conference}.\hskip 1em plus 0.5em minus 0.4em\relax
  IEEE, 2024, pp. 1--13.

\bibitem{pavur2019secrets}
J.~Pavur, D.~Moser, V.~Lenders, and I.~Martinovic, ``Secrets in the sky: on
  privacy and infrastructure security in dvb-s satellite broadband,'' in
  \emph{ACM Conference on Security and Privacy in Wireless and Mobile Networks
  (WiSec)}, 2019, pp. 277--284.

\bibitem{Pavur2021QPEPAnActionable}
\BIBentryALTinterwordspacing
J.~Pavur, M.~Strohmeier, V.~Lenders, and I.~Martinovic, ``Qpep: An actionable
  approach to secure and performant broadband from geostationary orbit,''
  \emph{Proceedings 2021 Network and Distributed System Security Symposium},
  2021. [Online]. Available:
  \url{https://api.semanticscholar.org/CorpusID:231879176}
\BIBentrySTDinterwordspacing

\bibitem{visionspaceYamcsVulns}
``{Y}amcs v5.8.6 {V}ulnerability {A}ssessment --- visionspace.com,''
  \url{https://visionspace.com/yamcs-v5-8-6-vulnerability-assessment/},
  [Accessed 28-07-2025].

\bibitem{honeysat}
\BIBentryALTinterwordspacing
E.~López-Morales, U.~Planta, G.~Marra, C.~González, J.~Hopkins, M.~Garoosi,
  E.~Obreque, C.~Rubio-Medrano, and A.~Abbasi, ``Honeysat: A network-based
  satellite honeypot framework,'' in \emph{Proceedings of the {NDSS}
  Symposium}, ser. NDSS '26.\hskip 1em plus 0.5em minus 0.4em\relax The
  Internet Society, 2026. [Online]. Available:
  \url{https://www.ndss-symposium.org/ndss-paper/honeysat-a-network-based-satellite-honeypot-framework/}
\BIBentrySTDinterwordspacing

\bibitem{salkield2023firefly}
E.~Salkield, S.~K{\"o}hler, S.~Birnbach, R.~Baker, M.~Strohmeier, and
  I.~Martinovic, ``Firefly: Spoofing earth observation satellite data through
  radio overshadowing,'' in \emph{Workshop on the Security of Space and
  Satellite Systems (SpaceSec)}, 2023.

\bibitem{wireless2024bisping}
\BIBentryALTinterwordspacing
R.~Bisping, J.~Willbold, M.~Strohmeier, and V.~Lenders, ``Wireless signal
  injection attacks on {VSAT} satellite modems,'' in \emph{USENIX Security
  Symposium}.\hskip 1em plus 0.5em minus 0.4em\relax Philadelphia, PA: USENIX
  Association, Aug. 2024, pp. 6075--6091. [Online]. Available:
  \url{https://www.usenix.org/conference/usenixsecurity24/presentation/bisping}
\BIBentrySTDinterwordspacing

\bibitem{saeidi2025enhancing}
T.~Saeidi and S.~Karamzadeh, ``Enhancing cubesat communication through
  beam-steering antennas: A review of technologies and challenges,''
  \emph{Electronics}, vol.~14, no.~4, p. 754, 2025.

\bibitem{aac_pulsar_advertisement_card}
{AAC Clyde Space}, ``Pulsar‑utrx uhf/uhf transceiver – high‑performance
  cubesat telemetry and command radio,''
  \url{https://www.aac-clyde.space/wp-content/uploads/2021/10/AAC\_DataSheet\_Pulsar\_UTRX-4.pdf},
  2023, compact PC/104‑form UHF telemetry and command radio, 9600 bps GMSK /
  1200 bps AFSK, 27–33 dBm TX power, SEU‑immune flash FPGA; release date
  June 6 2023.

\bibitem{sdls}
C.~S.~N. Aeronautics and S.~Administration, ``{S}{P}{A}{C}{E} {D}{A}{T}{A}
  {L}{I}{N}{K} {S}{E}{C}{U}{R}{I}{T}{Y} {P}{R}{O}{T}{O}{C}{O}{L},''
  \url{https://ccsds.org/Pubs/355x0b2.pdf}, [Accessed 23-06-2025].

\bibitem{planta2024let}
U.~Planta, J.~Rederlechner, G.~Marra, and A.~Abbasi, ``Let me do it for you: On
  the feasibility of inter-satellite friendly jamming,'' in \emph{2024 Security
  for Space Systems (3S)}.\hskip 1em plus 0.5em minus 0.4em\relax IEEE, 2024,
  pp. 1--6.

\bibitem{polysat_earthstation}
{PolySat Team}, ``{Earth Station – PolySat},''
  \url{https://www.polysat.org/earth-station}, 2025, accessed: 2025-07-03.

\bibitem{eoportalBreizhReconnaissance}
``{B}{R}{O} ({B}reizh {R}econnaissance {O}rbiter) / {U}nseenlabs - eo{P}ortal
  --- eoportal.org,''
  \url{https://www.eoportal.org/satellite-missions/unseenlabs\#references},
  [Accessed 22-07-2025].

\bibitem{eoportalKleosCommercial}
``{K}leos {C}ommercial {R}{F} {M}onitoring {C}ube{S}at {C}onstellation -
  eo{P}ortal --- eoportal.org,''
  \url{https://www.eoportal.org/satellite-missions/kleos\#mission-status},
  [Accessed 22-07-2025].

\bibitem{eoportalSpooQy1CubeSat}
``{S}poo{Q}y-1 {C}ube{S}at {M}ission - eo{P}ortal --- eoportal.org,''
  \url{https://www.eoportal.org/satellite-missions/spooqy-1\#spooqy-1-cubesat-mission-with-qkd-quantum-key-distribution},
  [Accessed 22-07-2025].

\bibitem{nanosatsERNSTNanosats}
E.~Kulu, ``{E}{R}{N}{S}{T} - {N}anosats {D}atabase --- nanosats.eu,''
  \url{https://www.nanosats.eu/sat/ernst}, [Accessed 22-07-2025].

\bibitem{skyrocketIONSCV002comma}
``{I}{O}{N}-{S}{C}{V} space.skyrocket.de,''
  \url{https://space.skyrocket.de/doc\_sdat/ion-scv-2.htm}, [Accessed
  22-07-2025].

\bibitem{marketsandmarketsCubeSatCompanies}
``{C}ube{S}at {C}ompanies - {T}op {C}ompanies {L}ist of {C}ube{S}at {I}ndustry
  --- marketsandmarkets.com,''
  \url{https://www.marketsandmarkets.com/ResearchInsight/cubesat-market.asp#:~:text=EnduroSat%20,solar%20panels%2C%20and%20custom%20modules},
  [Accessed 12-11-2025].

\bibitem{grandviewresearchCubeSatMarket}
``{C}ube{S}at {M}arket {S}ize, {S}hare \& {T}rends | {I}ndustry {R}eport, 2033
  --- grandviewresearch.com,''
  \url{https://www.grandviewresearch.com/industry-analysis/cubesat-market-report#:~:text=Key%20companies%20profiled},
  [Accessed 12-11-2025].

\bibitem{lucintelCubeSatMarket}
Lucintel, ``{C}ube{S}at {M}arket {R}eport: {T}rends, {F}orecast and
  {C}ompetitive {A}nalysis to 2030 --- lucintel.com,''
  \url{https://www.lucintel.com/cubesat-market.aspx}, [Accessed 13-11-2025].

\bibitem{fortunebusinessinsightsCubeSatMarket}
F.~B. Insights, ``{C}ube{S}at {M}arket {S}ize, {S}hare | {G}lobal {G}rowth
  {R}eport [2032] --- fortunebusinessinsights.com,''
  \url{https://www.fortunebusinessinsights.com/cubesat-market-113707}, 2025,
  [Accessed 13-11-2025].

\bibitem{imarcgroupCubeSatMarket}
I.~Group, ``{C}ube{S}at {M}arket {S}ize, {A}nalysis, {T}rends \& {F}orecast,
  2033 --- imarcgroup.com,'' \url{https://www.imarcgroup.com/cubesat-market},
  2024, [Accessed 13-11-2025].

\bibitem{endurosat_homepage}
{EnduroSat}, ``{EnduroSat – Satellites \& Space Missions},''
  \url{https://www.endurosat.com/}, accessed: 2025-07-15.

\bibitem{skyrocketSharedSat1comma}
``{S}hared{S} space.skyrocket.de,''
  \url{https://space.skyrocket.de/doc\_sdat/spartan.htm}, [Accessed
  22-07-2025].

\bibitem{curiumsatGitLab}
``curiumsat · {G}it{L}ab --- gitlab.com,'' \url{https://gitlab.com/curiumsat},
  [Accessed 22-07-2025].

\bibitem{fosdem2025satnogs}
``{F}{O}{S}{D}{E}{M} 2025 - {S}at{N}{O}{G}{S}-{C}{O}{M}{M}{S}: {A}n
  {O}pen-{S}ource {C}ommunication {S}ubsystem for {C}ube{S}ats ---
  fosdem.org,''
  \url{https://fosdem.org/2025/schedule/event/fosdem-2025-6024-satnogs-comms-an-open-source-communication-subsystem-for-cubesats/},
  [Accessed 28-07-2025].

\bibitem{freertosReleaseHistory}
``{F}ree{R}{T}{O}{S} {R}elease {H}istory - {F}ree{R}{T}{O}{S} ---
  freertos.org,''
  \url{https://freertos.org/Documentation/04-Roadmap-and-release-note/02-Release-notes/00-Release-history},
  [Accessed 22-07-2025].

\bibitem{LibcspLibcsp2024}
\BIBentryALTinterwordspacing
Y.~Shoji, ``libcsp/libcsp,'' Feb. 2024. [Online]. Available:
  \url{https://github.com/libcsp/libcsp}
\BIBentrySTDinterwordspacing

\bibitem{libcspComparisonLeaks}
``{M}{A}{C} comparison leaks timing data · {I}ssue \#44 · libcsp/libcsp ---
  github.com,'' \url{https://github.com/libcsp/libcsp/issues/44}, [Accessed
  22-07-2025].

\bibitem{camurati2018screaming}
G.~Camurati, S.~Poeplau, M.~Muench, T.~Hayes, and A.~Francillon, ``Screaming
  channels: When electromagnetic side channels meet radio transceivers,'' in
  \emph{ACM Conference on Computer and Communications Security (CCS)}, 2018,
  pp. 163--177.

\bibitem{gomspaceLinkSX}
``{G}{O}{M}space | {N}ano{C}om {L}ink {S}{X} --- gomspace.com,''
  \url{https://gomspace.com/shop/subsystems/communication-systems/nanocom-link-sx.aspx},
  [Accessed 22-07-2025].

\bibitem{satnogs_protobuf}
{Libre Space Foundation / SatNOGS Project}, ``Satnogs comms protobuf messaging
  specification,''
  \url{https://gitlab.com/librespacefoundation/satnogs-comms/satnogs-comms-proto},
  2023, open‑source protobuf definition for SatNOGS communications subsystem;
  created June 19, 2023.

\bibitem{palmer2025wireless}
B.~Palmer, V.~Schulz, M.~Jahnke, E.~Stoll, and U.~Kulau, ``Wireless
  intra-satellite lifi dual can bus networks for redundancy and throughput,''
  in \emph{Deutsche Luft-und Raumfahrtkongress 2025}, 2025.

\bibitem{2022HeRapidPatch}
\BIBentryALTinterwordspacing
Y.~He, Z.~Zou, K.~Sun, Z.~Liu, K.~Xu, Q.~Wang, C.~Shen, Z.~Wang, and Q.~Li,
  ``{RapidPatch}: Firmware hotpatching for {Real-Time} embedded devices,'' in
  \emph{USENIX Security Symposium}.\hskip 1em plus 0.5em minus 0.4em\relax
  Boston, MA: USENIX Association, Aug. 2022, pp. 2225--2242. [Online].
  Available:
  \url{https://www.usenix.org/conference/usenixsecurity22/presentation/he-yi}
\BIBentrySTDinterwordspacing

\end{thebibliography}
\appendices

\section{SatNOGS Survey Details}
\label{appendix:satnogs_details}
As mission operators rarely disclose details about their use of HMACs or encryption, we approximate this information based on Telemetry data recorded and released by the SatNOGS project \cite{satnogs}. Our approach filters the SatNOGS database for missions using our target transmitter. For each selected mission, we downloaded up to 10,000 telemetry samples. This limit was chosen to balance between obtaining sufficient data for meaningful analysis and minimizing the load on the SatNOGS API, especially since some missions have accumulated several million samples.
The downloaded data samples have undergone demodulation but not yet decoding. Consequently, we performed decoding ourselves based on the CSPv2 standard, which is used by the relevant transmitter.

Given our interest regarding the usage of HMACs within a mission, the most straightforward approach for processing these samples would be to examine the flags that have been set.
However, our investigations have indicated that these flags are not always utilized correctly. Therefore, we have decided to examine the actual HMACs to obtain results that are more aligned with the truth.

The processing pipeline is as follows:
\begin{enumerate}
    \item \textbf{Filtering Duplicate Samples:} Identical telemetry samples do not yield new insights about HMAC behavior, as their HMACs (if present) would remain unchanged. These duplicates are removed.
    \item \textbf{Filtering Incomplete Samples:}  We remove samples that are smaller than the CSP header size or have payloads too short to contain a truncated HMAC.
    \item \textbf{Filtering Encrypted Samples:} We exclude all samples with the encryption flag set, regardless of the HMAC flag status, since encrypted payloads prevent meaningful inspection of the HMAC.
    \item \textbf{Entropy Estimation via Hamming Distance:} Direct entropy analysis on 4-byte segments is infeasible due to the small size, so we approximate entropy using the Hamming distance between samples. While this method cannot distinguish whether a high Hamming distance is caused by high variance in the data or by the presence of an HMAC, it is reasonable to assume that the rarity of a large variance in the data transmitted will yield a limited number of outliers. Conversely, the implementation of an HMAC will result in consistent, substantially large Hamming distances in comparison.
\end{enumerate}
We then categorized missions based on the following criteria:
\begin{enumerate}
    \item \textbf{Not enough Data:} Missions with fewer than 20 usable samples are marked as having insufficient data for analysis.
    \item \textbf{Encryption Only:} If all but fewer than 5 samples have the encryption flag set, including the samples filtered for encryption earlier, we assume the dataset available is limited to only containing encrypted samples, with the exceptions potentially caused by bit flips or noise.
    \item \textbf{No HMAC:} A mission is tagged as not using HMACs if fewer than 5 usable samples have the HMAC flag set and the average Hamming distance across the presumed HMAC bytes is below 14. This threshold is chosen because the expected Hamming distance for truly random 4-byte sequences is 16; a value significantly below that suggests low entropy consistent with the absence of an HMAC.
\end{enumerate}
For all remaining missions that likely use HMACs, we identify three possible scenarios:
\begin{enumerate}
    \item \textbf{Consistent and Plausible HMAC usage:} The HMAC flag frequency aligns with the frequency of high Hamming distances in the presumed HMAC region, suggesting correct and consistent use of HMACs. For these missions, we further analyze whether they excursively use HMACs, or if a significant number of samples don't use HMACs. The latter suggests that the mission switches between using an HMAC and not using an HMAC.
    \item \textbf{Underused HMAC Flag:} If the HMAC flag count is significantly lower than the observed entropy, this may imply inconsistent flag usage or a large variation in the payload data that comes close to mimicking the randomness of an HMAC.
    \item \textbf{Overused HMAC Flag:} If the HMAC flag count is significantly higher than what the entropy suggests, this may indicate that the flag is used incorrectly or repurposed.
\end{enumerate}
For the latter two scenarios, we cannot confidently determine the true HMAC usage.

 \section{Additional Data/Attack Flows}
\label{appendix:taxonomy}
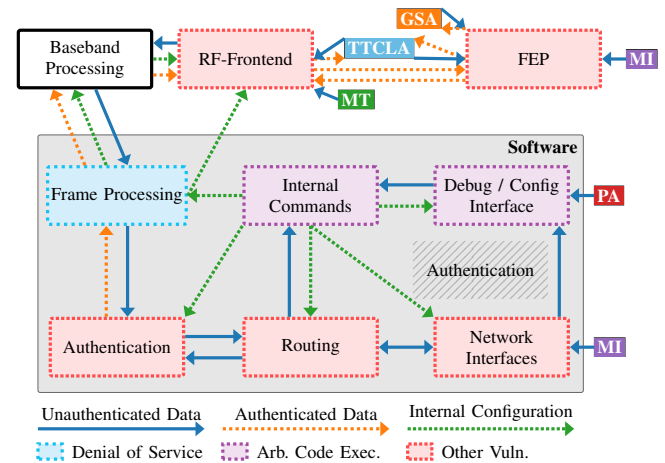
\begin{figure}[H]
  \centering
  \resizebox{1.0\columnwidth}{!}{
    \begin{tikzpicture}[
  font=\LARGE,
  box/.style={draw, rounded corners=2pt, minimum width=4.4cm, minimum height=1.9cm, align=center, line width=3pt},
  colsep/.store in=\colsep, colsep=1.8cm,   rowsep/.store in=\rowsep, rowsep=3.0cm,    vulnbox/.style={draw=red!70, rounded corners=2pt, minimum width=4.4cm, minimum height=1.9cm, align=center, line width=3pt, fill=red!12, dotted},
  rcebox/.style={draw=violet!70, rounded corners=2pt, minimum width=4.4cm, minimum height=1.9cm, align=center, line width=3pt, fill=violet!12, dashed},
  dosbox/.style={draw=cyan!70, rounded corners=2pt, minimum width=4.4cm, minimum height=1.9cm, align=center, line width=3pt, fill=cyan!15, loosely dash dot},
  missing/.style={
  draw=none, minimum width=4.4cm, minimum height=1.9cm,
  pattern=north east lines,
  pattern color=gray!60
}
]

\definecolor{SigAColor}{HTML}{1F77B4} \definecolor{SigBColor}{HTML}{FF7F0E} \definecolor{SigCColor}{HTML}{2CA02C}

\tikzset{
  siguauth/.style={->, draw=SigAColor, line width=3.0pt, >={Triangle[length=3mm]}},             sigauth/.style={->, draw=SigBColor, line width=3.0pt, dashed, >={Triangle[length=3mm]}},sigconf/.style={->, draw=SigCColor, line width=3.0pt, dotted, >={Triangle[length=3mm]}}      }
\coordinate (frameNW) at (-9.0,2.0);
\coordinate (frameSE) at (9,-6.5);
\draw[draw=black!60, thick, fill=black!10, rounded corners=2pt] (frameNW) rectangle (frameSE);
\node[anchor=north east] at ([xshift=-2pt,yshift=-2pt]frameNE) {\textbf{Software}};
\node[rcebox] (intcmd) {Internal\\Commands};
\node[dosbox, left=\colsep of intcmd] (rfproc) {Frame Processing};
\node[rcebox, right=\colsep of intcmd] (dbg) {Debug / Config\\Interface};

\node[vulnbox, below=\rowsep of rfproc] (auth) {Authentication};
\node[missing, xshift= -20pt, below=0.5cm of dbg] (intauth) {Authentication};
\node[vulnbox, below=\rowsep of intcmd] (route) {Routing};
\node[vulnbox, below=\rowsep of dbg] (netif) {Network\\Interfaces};

\node[box, above=2.5cm of rfproc, xshift= -30pt] (bb) {Baseband\\Processing};
\node[vulnbox, right=0.8cm of bb] (rf) {RF-Frontend};

\coordinate (frameSW) at (frameNW |- frameSE);
\coordinate (frameNE) at (frameSE |- frameNW);

\node[draw, minimum width=0.55cm, minimum height=0.45cm, inner sep=0pt,
      right=0.8cm of dbg] (pTop) {\LARGE\attackerPA{}};
\node[draw, minimum width=0.55cm, minimum height=0.45cm, inner sep=0pt,
      right=0.8cm of netif] (pBot) {\LARGE\attackerMI{}};

\node[vulnbox, right=5.0cm of rf] (com) {FEP};

\node[draw, minimum width=0.55cm, minimum height=0.45cm, inner sep=0pt,
      right=0.8cm of com] (pFep) {\LARGE\attackerMI{}};

\node[draw, minimum width=0.55cm, minimum height=0.45cm, inner sep=0pt, yshift=-38pt,
      right=0.8cm of rf] (pCom) {\LARGE\attackerMT{}};

\node[draw, minimum width=0.55cm, minimum height=0.45cm, inner sep=0pt, yshift=38pt,
      left=0.8cm of com] (pRF) {\LARGE\attackerGSA{}};

\node[draw, minimum width=0.55cm, minimum height=0.45cm, inner sep=0pt,
      left=1.7cm of com, yshift=10pt] (pMiddle) {\LARGE\attackerAitM{}};

\draw[siguauth, transform canvas={yshift= 15pt}] (rf.west) -- (bb.east);
\draw[sigauth, transform canvas={yshift= -15pt}] (bb.east) -- (rf.west);
\draw[sigconf] (bb.east) -- (rf.west);

\draw[sigconf] (rfproc.east) -- (rf.south);

\draw[siguauth, transform canvas={xshift= 10pt}] (bb.south) -- (rfproc.north);
\draw[sigconf, transform canvas={xshift= -10pt}] (rfproc.north) -- (bb.south);
\draw[sigauth, transform canvas={xshift= -30pt}] (rfproc.north) -- (bb.south);

\draw[siguauth] (pTop.west) -- (dbg.east);

\draw[siguauth, transform canvas={xshift= 10pt}] (rfproc.south) -- (auth.north);
\draw[sigauth, transform canvas={xshift= -10pt}] (auth.north) -- (rfproc.south);

\draw[siguauth, transform canvas={yshift= 10pt}] (auth.east) -- (route.west);
\draw[siguauth, transform canvas={yshift= -10pt}] (route.west) -- (auth.east);

\draw[<->, draw=SigAColor, line width=3pt, >={Triangle[length=3mm]}] (route.east) -- (netif.west);

\draw[siguauth, transform canvas={yshift= 10pt}] (dbg.west) -- (intcmd.east);
\draw[sigconf, transform canvas={yshift= -10pt}] (intcmd.east) -- (dbg.west);

\draw[siguauth] (pBot.west) -- (netif.east);

\draw[siguauth, line width=1mm] (pRF.north east) -- (com.north west);
\draw[sigauth, line width=1mm] (com.north west) -- (pRF.south east);
\draw[siguauth, line width=1mm] (pCom.west) -- (rf.south east);

\draw[sigauth, line width=1mm, transform canvas={yshift= -10pt}] (rf.east) -- (com.west);

\draw[sigauth, line width=1mm, transform canvas={yshift= -20pt}] (com.west) -- (rf.east);

\draw[sigauth] (com.west) -- (pMiddle.north east);
\draw[siguauth] (pMiddle.south east) -- (com.west);

\draw[sigauth] (rf.east) -- (pMiddle.south west);
\draw[siguauth] (pMiddle.north west) -- (rf.east);

\draw[siguauth] (pFep.west) -- (com.east);

\draw[sigconf] (intcmd.south) -- (route.north);
\draw[siguauth, transform canvas={xshift= -20pt}] (route.north) -- (intcmd.south);

\draw[sigconf] (intcmd.west) -- (rfproc.east);

\draw[sigconf] (intcmd.south west) -- (auth.north east);

\draw[sigconf] (intcmd.south) -- (netif.north west);

\draw[siguauth, transform canvas={xshift= -10pt}] (netif.north east) -- (dbg.south east);

\def\L{55mm}  \def\S{6mm}

\coordinate (legendOrigin) at ($ (frameSW)!0.5!(frameSE) + (-90mm,-12mm) $);

\coordinate (LA) at (legendOrigin);
\coordinate (LB) at ($(legendOrigin)+(\L+\S,0)$);
\coordinate (LC) at ($(legendOrigin)+(2*\L+2*\S,0)$);

\draw[siguauth] (LA) -- ++(\L,0) node[midway,above=1mm]{Unauthenticated Data};
\draw[sigauth] (LB) -- ++(\L,0) node[midway,above=1mm]{Authenticated Data};
\draw[sigconf] (LC) -- ++(\L,0) node[midway,above=1mm]{Internal Configuration};

\def\BW{8mm}   \def\BH{5mm}   \def\Yoff{10mm}

\coordinate (L2A) at ($(LA)+(0,-\Yoff)$);
\coordinate (L2B) at ($(LB)+(0,-\Yoff)$);
\coordinate (L2C) at ($(LC)+(0,-\Yoff)$);

\path[draw=cyan!70, line width=3pt, fill=cyan!15,loosely dash dot]
  (L2A) rectangle ++(\BW,\BH);
\node[anchor=west] at ($(L2A)+(\BW+2mm,0.5*\BH)$) {Denial of Service};

\path[draw=violet!70, line width=3pt, dashed, fill=violet!12]
  (L2B) rectangle ++(\BW,\BH);
\node[anchor=west] at ($(L2B)+(\BW+2mm,0.5*\BH)$) {Arb. Code Exec.};

\path[draw=red!70, line width=3pt, dotted, fill=red!12]
  (L2C) rectangle ++(\BW,\BH);
\node[anchor=west] at ($(L2C)+(\BW+2mm,0.5*\BH)$) {Other Vuln.};

\end{tikzpicture}

  }
  \caption{Sub-component based view of data/attack flow for the \GomSpace{} \AXonehundred{}.}
  \label{fig:taxonomy_gom}
\end{figure}
 \begin{figure}[H]
  \centering
  \resizebox{1.0\columnwidth}{!}{
    \begin{tikzpicture}[
  font=\LARGE,
  box/.style={draw, rounded corners=2pt, minimum width=4.4cm, minimum height=1.9cm, align=center, line width=3pt},
  colsep/.store in=\colsep, colsep=1.8cm,   rowsep/.store in=\rowsep, rowsep=3.0cm,    vulnbox/.style={draw=red!70, rounded corners=2pt, minimum width=4.4cm, minimum height=1.9cm, align=center, line width=3pt, fill=red!12, dotted},
  rcebox/.style={draw=violet!70, rounded corners=2pt, minimum width=4.4cm, minimum height=1.9cm, align=center, line width=3pt, fill=violet!12, dashed},
  dosbox/.style={draw=cyan!70, rounded corners=2pt, minimum width=4.4cm, minimum height=1.9cm, align=center, line width=3pt, fill=cyan!15},
  missing/.style={
  draw=none, minimum width=4.4cm, minimum height=1.9cm,
  pattern=north east lines,
  pattern color=gray!60
}
]

\definecolor{SigAColor}{HTML}{1F77B4} \definecolor{SigBColor}{HTML}{FF7F0E} \definecolor{SigCColor}{HTML}{2CA02C}

\tikzset{
  siguauth/.style={->, draw=SigAColor, line width=3.0pt, >={Triangle[length=3mm]}},             sigauth/.style={->, draw=SigBColor, line width=3.0pt, dashed, >={Triangle[length=3mm]}},sigconf/.style={->, draw=SigCColor, line width=3.0pt, dotted, >={Triangle[length=3mm]}}      }
\coordinate (frameNW) at (-9.0,2.0);
\coordinate (frameSE) at (9,-6.5);
\draw[draw=black!60, thick, fill=black!10, rounded corners=2pt] (frameNW) rectangle (frameSE);
\node[anchor=north east] at ([xshift=-2pt,yshift=-2pt]frameNE) {\textbf{Software}};
\node[rcebox] (intcmd) {Internal\\Commands};
\node[box, left=\colsep of intcmd] (rfproc) {Frame Processing};
\node[rcebox, right=\colsep of intcmd] (dbg) {Debug / Config\\Interface};

\node[box, below=\rowsep of rfproc] (auth) {Authentication};
\node[missing, xshift= -20pt, below=0.5cm of dbg] (intauth) {Authentication};
\node[box, below=\rowsep of intcmd] (route) {Routing};
\node[box, below=\rowsep of dbg] (netif) {Network\\Interfaces};

\node[box, above=2.5cm of rfproc, xshift= -30pt] (bb) {Baseband\\Processing};
\node[box, right=0.8cm of bb] (rf) {RF-Frontend};

\coordinate (frameSW) at (frameNW |- frameSE);
\coordinate (frameNE) at (frameSE |- frameNW);

\node[draw, minimum width=0.55cm, minimum height=0.45cm, inner sep=0pt,
      right=0.8cm of dbg] (pTop) {\LARGE\attackerPA{}};
\node[draw, minimum width=0.55cm, minimum height=0.45cm, inner sep=0pt,
      right=0.8cm of netif] (pBot) {\LARGE\attackerMI{}};

\node[box, right=5.0cm of rf] (com) {FEP};

\node[draw, minimum width=0.55cm, minimum height=0.45cm, inner sep=0pt,
      right=0.8cm of com] (pFep) {\LARGE\attackerMI{}};

\node[draw, minimum width=0.55cm, minimum height=0.45cm, inner sep=0pt, yshift=-38pt,
      right=0.8cm of rf] (pCom) {\LARGE\attackerMT{}};

\node[draw, minimum width=0.55cm, minimum height=0.45cm, inner sep=0pt, yshift=38pt,
      left=0.8cm of com] (pRF) {\LARGE\attackerGSA{}};

\node[draw, minimum width=0.55cm, minimum height=0.45cm, inner sep=0pt,
      left=1.7cm of com, yshift=10pt] (pMiddle) {\LARGE\attackerAitM{}};

\draw[siguauth, transform canvas={yshift= 15pt}] (rf.west) -- (bb.east);
\draw[sigauth, transform canvas={yshift= -15pt}] (bb.east) -- (rf.west);
\draw[sigconf] (bb.east) -- (rf.west);

\draw[sigconf] (rfproc.east) -- (rf.south);

\draw[siguauth, transform canvas={xshift= 10pt}] (bb.south) -- (rfproc.north);
\draw[sigconf, transform canvas={xshift= -10pt}] (rfproc.north) -- (bb.south);
\draw[sigauth, transform canvas={xshift= -30pt}] (rfproc.north) -- (bb.south);

\draw[siguauth] (pTop.west) -- (dbg.east);

\draw[siguauth, transform canvas={xshift= 10pt}] (rfproc.south) -- (auth.north);
\draw[sigauth, transform canvas={xshift= -10pt}] (auth.north) -- (rfproc.south);

\draw[sigauth, transform canvas={yshift= 10pt}] (auth.east) -- (route.west);
\draw[siguauth, transform canvas={yshift= -10pt}] (route.west) -- (auth.east);

\draw[<->, draw=SigAColor, line width=3pt, >={Triangle[length=3mm]}] (route.east) -- (netif.west);

\draw[siguauth] (pBot.west) -- (netif.east);

\draw[siguauth, line width=1mm] (pRF.north east) -- (com.north west);
\draw[sigauth, line width=1mm] (com.north west) -- (pRF.south east);
\draw[siguauth, line width=1mm] (pCom.west) -- (rf.south east);

\draw[sigauth, line width=1mm, transform canvas={yshift= -10pt}] (rf.east) -- (com.west);

\draw[sigauth, line width=1mm, transform canvas={yshift= -20pt}] (com.west) -- (rf.east);

\draw[sigauth] (com.west) -- (pMiddle.north east);
\draw[siguauth] (pMiddle.south east) -- (com.west);

\draw[sigauth] (rf.east) -- (pMiddle.south west);
\draw[siguauth] (pMiddle.north west) -- (rf.east);

\draw[siguauth] (pFep.west) -- (com.east);

\draw[sigconf] (intcmd.south) -- (route.north);
\draw[siguauth, transform canvas={xshift= -20pt}] (route.north) -- (intcmd.south);

\draw[sigconf] (intcmd.west) -- (rfproc.east);

\draw[sigconf] (intcmd.south west) -- (auth.north east);

\draw[sigconf] (intcmd.south) -- (netif.north west);

\def\L{55mm}  \def\S{6mm}

\coordinate (legendOrigin) at ($ (frameSW)!0.5!(frameSE) + (-90mm,-12mm) $);

\coordinate (LA) at (legendOrigin);
\coordinate (LB) at ($(legendOrigin)+(\L+\S,0)$);
\coordinate (LC) at ($(legendOrigin)+(2*\L+2*\S,0)$);

\draw[siguauth] (LA) -- ++(\L,0) node[midway,above=1mm]{Unauthenticated Data};
\draw[sigauth] (LB) -- ++(\L,0) node[midway,above=1mm]{Authenticated Data};
\draw[sigconf] (LC) -- ++(\L,0) node[midway,above=1mm]{Internal Configuration};

\def\BW{8mm}   \def\BH{5mm}   \def\Yoff{10mm}

\coordinate (L2A) at ($(LA)+(0,-\Yoff)$);
\coordinate (L2B) at ($(LB)+(0,-\Yoff)$);
\coordinate (L2C) at ($(LC)+(0,-\Yoff)$);

\path[draw=cyan!70, line width=3pt, fill=cyan!15, loosely dash dot]
  (L2A) rectangle ++(\BW,\BH);
\node[anchor=west] at ($(L2A)+(\BW+2mm,0.5*\BH)$) {Denial of Service};

\path[draw=violet!70, line width=3pt, dashed, fill=violet!12]
  (L2B) rectangle ++(\BW,\BH);
\node[anchor=west] at ($(L2B)+(\BW+2mm,0.5*\BH)$) {Arb. Code Exec.};

\path[draw=red!70, line width=3pt, dotted, fill=red!12]
  (L2C) rectangle ++(\BW,\BH);
\node[anchor=west] at ($(L2C)+(\BW+2mm,0.5*\BH)$) {Other Vuln.};

\end{tikzpicture}
  }
  \label{fig:taxonomy_enduro}
\end{figure}
 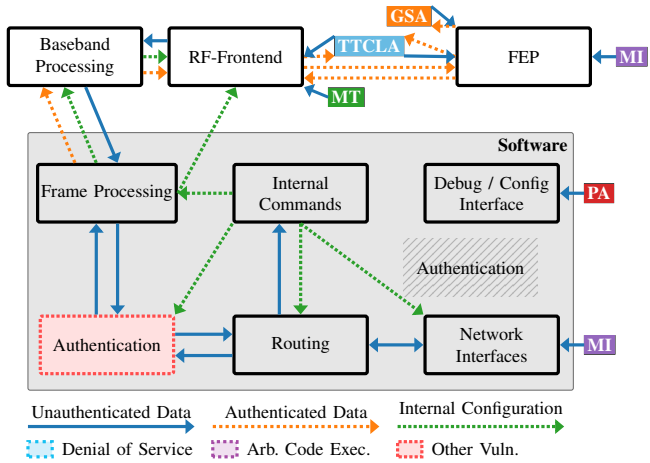
\begin{figure}[H]
  \centering
  \resizebox{1.0\columnwidth}{!}{
    \begin{tikzpicture}[
  font=\LARGE,
  box/.style={draw, rounded corners=2pt, minimum width=4.4cm, minimum height=1.9cm, align=center, line width=3pt},
  colsep/.store in=\colsep, colsep=1.8cm,   rowsep/.store in=\rowsep, rowsep=3.0cm,    vulnbox/.style={draw=red!70, rounded corners=2pt, minimum width=4.4cm, minimum height=1.9cm, align=center, line width=3pt, fill=red!12, dotted},
  rcebox/.style={draw=violet!70, rounded corners=2pt, minimum width=4.4cm, minimum height=1.9cm, align=center, line width=3pt, fill=violet!12, dashed},
  dosbox/.style={draw=cyan!70, rounded corners=2pt, minimum width=4.4cm, minimum height=1.9cm, align=center, line width=3pt, fill=cyan!15},
  missing/.style={
  draw=none, minimum width=4.4cm, minimum height=1.9cm,
  pattern=north east lines,
  pattern color=gray!60
}
]

\definecolor{SigAColor}{HTML}{1F77B4} \definecolor{SigBColor}{HTML}{FF7F0E} \definecolor{SigCColor}{HTML}{2CA02C}

\tikzset{
  siguauth/.style={->, draw=SigAColor, line width=3.0pt, >={Triangle[length=3mm]}},             sigauth/.style={->, draw=SigBColor, line width=3.0pt, dashed, >={Triangle[length=3mm]}},sigconf/.style={->, draw=SigCColor, line width=3.0pt, dotted, >={Triangle[length=3mm]}}      }
\coordinate (frameNW) at (-9.0,2.0);
\coordinate (frameSE) at (9,-6.5);
\draw[draw=black!60, thick, fill=black!10, rounded corners=2pt] (frameNW) rectangle (frameSE);
\node[anchor=north east] at ([xshift=-2pt,yshift=-2pt]frameNE) {\textbf{Software}};
\node[box] (intcmd) {Internal\\Commands};
\node[box, left=\colsep of intcmd] (rfproc) {Frame Processing};
\node[box, right=\colsep of intcmd] (dbg) {Debug / Config\\Interface};

\node[vulnbox, below=\rowsep of rfproc] (auth) {Authentication};
\node[missing, xshift= -20pt, below=0.5cm of dbg] (intauth) {Authentication};
\node[box, below=\rowsep of intcmd] (route) {Routing};
\node[box, below=\rowsep of dbg] (netif) {Network\\Interfaces};

\node[box, above=2.5cm of rfproc, xshift= -30pt] (bb) {Baseband\\Processing};
\node[box, right=0.8cm of bb] (rf) {RF-Frontend};

\coordinate (frameSW) at (frameNW |- frameSE);
\coordinate (frameNE) at (frameSE |- frameNW);

\node[draw, minimum width=0.55cm, minimum height=0.45cm, inner sep=0pt,
      right=0.8cm of dbg] (pTop) {\LARGE\attackerPA{}};
\node[draw, minimum width=0.55cm, minimum height=0.45cm, inner sep=0pt,
      right=0.8cm of netif] (pBot) {\LARGE\attackerMI{}};

\node[box, right=5.0cm of rf] (com) {FEP};

\node[draw, minimum width=0.55cm, minimum height=0.45cm, inner sep=0pt,
      right=0.8cm of com] (pFep) {\LARGE\attackerMI{}};

\node[draw, minimum width=0.55cm, minimum height=0.45cm, inner sep=0pt, yshift=-38pt,
      right=0.8cm of rf] (pCom) {\LARGE\attackerMT{}};

\node[draw, minimum width=0.55cm, minimum height=0.45cm, inner sep=0pt, yshift=38pt,
      left=0.8cm of com] (pRF) {\LARGE\attackerGSA{}};

\node[draw, minimum width=0.55cm, minimum height=0.45cm, inner sep=0pt,
      left=1.7cm of com, yshift=10pt] (pMiddle) {\LARGE\attackerAitM{}};

\draw[siguauth, transform canvas={yshift= 15pt}] (rf.west) -- (bb.east);
\draw[sigauth, transform canvas={yshift= -15pt}] (bb.east) -- (rf.west);
\draw[sigconf] (bb.east) -- (rf.west);

\draw[sigconf] (rfproc.east) -- (rf.south);

\draw[siguauth, transform canvas={xshift= 10pt}] (bb.south) -- (rfproc.north);
\draw[sigconf, transform canvas={xshift= -10pt}] (rfproc.north) -- (bb.south);
\draw[sigauth, transform canvas={xshift= -30pt}] (rfproc.north) -- (bb.south);

\draw[siguauth] (pTop.west) -- (dbg.east);

\draw[siguauth, transform canvas={xshift= 10pt}] (rfproc.south) -- (auth.north);
\draw[siguauth, transform canvas={xshift= -10pt}] (auth.north) -- (rfproc.south);

\draw[siguauth, transform canvas={yshift= 10pt}] (auth.east) -- (route.west);
\draw[siguauth, transform canvas={yshift= -10pt}] (route.west) -- (auth.east);

\draw[<->, draw=SigAColor, line width=3pt, >={Triangle[length=3mm]}] (route.east) -- (netif.west);

\draw[siguauth] (pBot.west) -- (netif.east);

\draw[siguauth, line width=1mm] (pRF.north east) -- (com.north west);
\draw[sigauth, line width=1mm] (com.north west) -- (pRF.south east);
\draw[siguauth, line width=1mm] (pCom.west) -- (rf.south east);

\draw[sigauth, line width=1mm, transform canvas={yshift= -10pt}] (rf.east) -- (com.west);

\draw[sigauth, line width=1mm, transform canvas={yshift= -20pt}] (com.west) -- (rf.east);

\draw[sigauth] (com.west) -- (pMiddle.north east);
\draw[siguauth] (pMiddle.south east) -- (com.west);

\draw[sigauth] (rf.east) -- (pMiddle.south west);
\draw[siguauth] (pMiddle.north west) -- (rf.east);

\draw[siguauth] (pFep.west) -- (com.east);

\draw[sigconf] (intcmd.south) -- (route.north);
\draw[siguauth, transform canvas={xshift= -20pt}] (route.north) -- (intcmd.south);

\draw[sigconf] (intcmd.west) -- (rfproc.east);

\draw[sigconf] (intcmd.south west) -- (auth.north east);

\draw[sigconf] (intcmd.south) -- (netif.north west);

\def\L{55mm}  \def\S{6mm}

\coordinate (legendOrigin) at ($ (frameSW)!0.5!(frameSE) + (-90mm,-12mm) $);

\coordinate (LA) at (legendOrigin);
\coordinate (LB) at ($(legendOrigin)+(\L+\S,0)$);
\coordinate (LC) at ($(legendOrigin)+(2*\L+2*\S,0)$);

\draw[siguauth] (LA) -- ++(\L,0) node[midway,above=1mm]{Unauthenticated Data};
\draw[sigauth] (LB) -- ++(\L,0) node[midway,above=1mm]{Authenticated Data};
\draw[sigconf] (LC) -- ++(\L,0) node[midway,above=1mm]{Internal Configuration};

\def\BW{8mm}   \def\BH{5mm}   \def\Yoff{10mm}

\coordinate (L2A) at ($(LA)+(0,-\Yoff)$);
\coordinate (L2B) at ($(LB)+(0,-\Yoff)$);
\coordinate (L2C) at ($(LC)+(0,-\Yoff)$);

\path[draw=cyan!70, line width=3pt, fill=cyan!15, loosely dash dot]
  (L2A) rectangle ++(\BW,\BH);
\node[anchor=west] at ($(L2A)+(\BW+2mm,0.5*\BH)$) {Denial of Service};

\path[draw=violet!70, line width=3pt, dashed, fill=violet!12]
  (L2B) rectangle ++(\BW,\BH);
\node[anchor=west] at ($(L2B)+(\BW+2mm,0.5*\BH)$) {Arb. Code Exec.};

\path[draw=red!70, line width=3pt, dotted, fill=red!12]
  (L2C) rectangle ++(\BW,\BH);
\node[anchor=west] at ($(L2C)+(\BW+2mm,0.5*\BH)$) {Other Vuln.};

\end{tikzpicture}

  }
  \caption{Sub-component based view of data/attack flow for the \LibreSpace{} \NogsComm{}.Note that the MCU debug interface being open is at digression of the mission owners and thus cannot be counted as a flaw of the \com{} product}
  \label{fig:taxonomy_nogs}
\end{figure}
  \section{RF-Measurements of AX100}
\begin{figure}[H] \centering
\includegraphics[width=\linewidth]{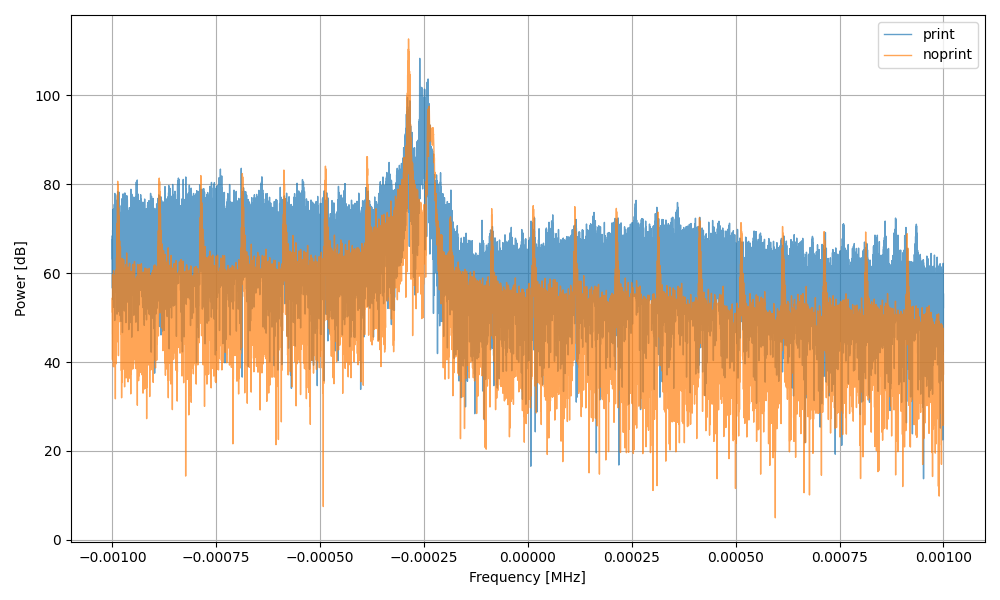}
  \caption{Measured RF spectrum of the transceiver while the MCU executes
           different workloads.}
  \label{fig:rf_spectrum_workloads}
\end{figure}
\newpage
\section{Meta-Review}

The following meta-review was prepared by the program committee for the 2026
IEEE Symposium on Security and Privacy (S\&P) as part of the review process as
detailed in the call for papers.

\subsection{Summary}
This paper analyzes the security of commercial off-the-shelf (COTS) satellite communication (COM) subsystems used in small satellites. The authors examine representative COM modules from several vendors using reverse engineering and testing techniques, identify vulnerabilities in firmware and system design, and discuss potential implications for deployed satellite missions. The work aims to connect weaknesses in real-world implementations with practical attack scenarios against satellite communication systems.

\subsection{Scientific Contributions}
\begin{itemize}
\item Identifies an Impactful Vulnerability.
\item Provides a Valuable Step Forward in an Established Field.
\item Establishes a New Research Direction.
\end{itemize}

\subsection{Reasons for Acceptance}
\begin{enumerate}
\item The paper presents an empirical security analysis of widely used COTS satellite communication modules and identifies vulnerabilities in real implementations.
\item The work highlights satellite COM modules as an important security boundary and reveals weaknesses in widely reused components that may affect multiple deployments.
\end{enumerate}

\end{document}